\documentclass[runningheads,a4paper]{llncs}

\usepackage{amssymb}
\usepackage{graphicx}
\usepackage{adjustbox}
\usepackage{array}
\usepackage{caption}
\usepackage{multirow}
\usepackage{float}
\usepackage{tabulary}
\usepackage{caption}
\usepackage[caption = false]{subfig}
\usepackage{graphicx}
\usepackage{url}
\usepackage{array}
\usepackage[colorinlistoftodos,prependcaption,textsize=small]{todonotes}
\usepackage{url}
\usepackage{breakurl} 

\newcommand{\keywords}[1]{\par\addvspace\baselineskip
\noindent\keywordname\enspace\ignorespaces#1}

\usepackage{natbib}
\setcitestyle{authoryear,open={(},close={)}}

\begin{document}

\mainmatter  

\title{Digital Forensics for IoT and WSNs}

\titlerunning{Lecture Notes in Computer Science: Authors' Instructions}

\author{Umit Karabiyik\inst{1}\thanks{Corresponding author} \and Kemal Akkaya\inst{2}}
\authorrunning{Lecture Notes in Computer Science: Authors' Instructions}


\institute{Purdue University\\ 
	Department of Computer and Information Technology\\
	West Lafayette IN 47906, USA\\
	\email{ukarabiy@purdue.edu}\\
\and
Florida International University\\ 
Department of Electrical and Computer Engineering\\
Miami FL 33174, USA\\
\email{kakkaya@fiu.edu}}

\toctitle{Lecture Notes in Computer Science}
\tocauthor{Authors' Instructions}
\maketitle

\renewcommand{\thefootnote}{\arabic{footnote}}

 \begin{abstract}
 In the last decade, wireless sensor networks (WSNs) and Internet-of-Things (IoT) devices are proliferated in many domains including critical infrastructures such as energy, transportation and manufacturing. Consequently, most of the daily operations now rely on the data coming from wireless sensors or IoT devices and their actions. In addition, personal IoT devices are heavily used for social media applications, which connect people as well as all critical infrastructures to each other under the cyber domain. However, this connectedness also comes with the risk of increasing number of cyber attacks through WSNs and/or IoT. While a significant research has been dedicated to secure WSN/IoT, this still indicates that there needs to be forensics mechanisms to be able to conduct investigations and analysis. In particular, understanding what has happened after a failure is crucial to many businesses, which rely on WSN/IoT applications. Therefore, there is a great interest and need for understanding digital forensics applications in WSN and IoT realms. This chapter fills this gap by providing an overview and classification of digital forensics research and applications in these emerging domains in a comprehensive manner. In addition to analyzing the technical challenges, the chapter provides a survey of the existing efforts from the device level to network level while also pointing out future research opportunities. 
 
 \keywords{digital forensics, IoT, WSN, security}
 \end{abstract}

\section{Introduction}
Wireless Sensor Networks (WSNs) have been initially proposed for military operations by the end of 90s \citep{estrin1999next}. However, with their potential in many applications, they have started to be deployed in different civil applications in early 2000s. WSNs have been touted to be used in many applications. These include but is not limited to environmental monitoring, habitat monitoring, structural health monitoring, health applications, agriculture applications and surveillance \citep{xu2002survey}. Typically, in such applications, a large number of sensors are deployed to sense the environment and send the collected data to a gateway or base-station for further processing. The communication is multi-hop and all the nodes are assumed to be battery-operated with limited processing and storage capabilities. There has always been incredible interest in WSN research from node level to application level \citep{akyildiz2002survey}.  The bulk of WSN research has focused on energy-efficient protocols at different layers of the protocol stack. The goal was to maximize the lifetime of the WSNs while enabling distributed operations. Energy-efficient MAC, routing, and transport protocols have been proposed \citep{akkaya2005survey,demirkol2006mac,wang2006survey}. Later, these protocols were augmented with security capabilities \citep{walters2007wireless}. Despite the huge amount of research, the WSN market was not mature. In the early 2000s, there were only a few sensor products (such as Mica2) and standardization efforts were not adequate. Therefore, the use of term WSN has been diminished and efforts are directed towards more personal sensor devices that came with the proliferation of smart phones and other wearable devices. The attention has shifted to these devices, referred to as Internet of Things (IoT). 

The term IoT  was first phrased in the context of supply chain management by Kevin Ashton in 1999 to get executive attention at Procter \& Gamble \citep{ashton2009internet}. Although it was used in different and somewhat related concepts earlier, the definition has become more comprehensive to comprise devices from health-care to entertainment and transportation to building management \citep{sundmaeker2010vision}. Therefore, the term might be used to describe the world where other devices are uniquely distinguishable, addressable, and contactable by means of the Internet. For example smart homes are furnished with hi-tech devices controlling such devices as the TV, refrigerator, microwave, blinds, music system, air conditioning units.

Today, we have more than 5 billion ``things" connected to the Internet and this number is expected to be nearly 50 billion (there are also different estimates) by 2020 \citep{iot2020}. Taking the advantage of using RFID and sensor network technology, physical objects such as computers, phones, wearable technologies, home appliances, vehicles, medical devices and industrial systems can be easily connected, tracked and managed by a single system \citep{jiang2014iot}. One of the many reasons to get these devices connected is that most people all want to take advantage of being conveniently ``online" in this age of Internet. On the other hand, we do underestimate the down sides of being connected in every second of every day.   

Although the expected number of connected devices is hypothetical, there is a real issue regarding existence of such a large collection of devices which are mostly vulnerable to cyberattacks. On October 21, 2016, we faced the reality of how our innocent household devices connected to the Internet could be part of an IoT army committing distributed denial of service attack (DDoS) to shut down websites including Twitter, Netflix, Paypal and Amazon Web services \citep{iotArmy}. In addition to being vulnerable, Syed Zaeem Hosain, CTO of Aeris - a pioneer in the machine-to-machine market, has raised the concern that scalability in IoT is the biggest issue as such a large number of devices will be generating enormously big data \citep{iotreality}. Hence, the following questions are asked by Mr. Hosain:

\begin{itemize}
	\item How will we transport such large data? 
	\item How will we store it? 
	\item How will we analyze it? 
	\item How will we search/find targeted data in large collection? 
	\item How will we keep the data secure and private? 	
\end{itemize}

All of these questions are part of our concerns about IoT today, however it is urgent that these issues must be addressed in advance before we are faced with serious scalability issues.

Miorandi et al. have discussed that IoT is a leading technology which brings various areas from cyber and the physical world together by the means of making physical devices smarter and connected with one another \citep{miorandi2012internet}. By taking this into account, the usage of the term IoT can be generalized into the following broad areas as discussed in \citep{atzori2010internet, pena2005itu}:

\begin{itemize}
  \item the global network providing ultimate interconnection ability to the smart things (devices) via the Internet 
  \item the collection of assistive technologies (e.g. RFIDs, Near Field Communication devices, and WISP.)
  \item the group of applications and services (e.g. Cloud services and Web of things.)
\end{itemize}

\begin{figure}
	\subfloat[Search trends for IoT, Wireless Sensor Networks and Ubiquitous Computing]{\includegraphics[width = \textwidth]{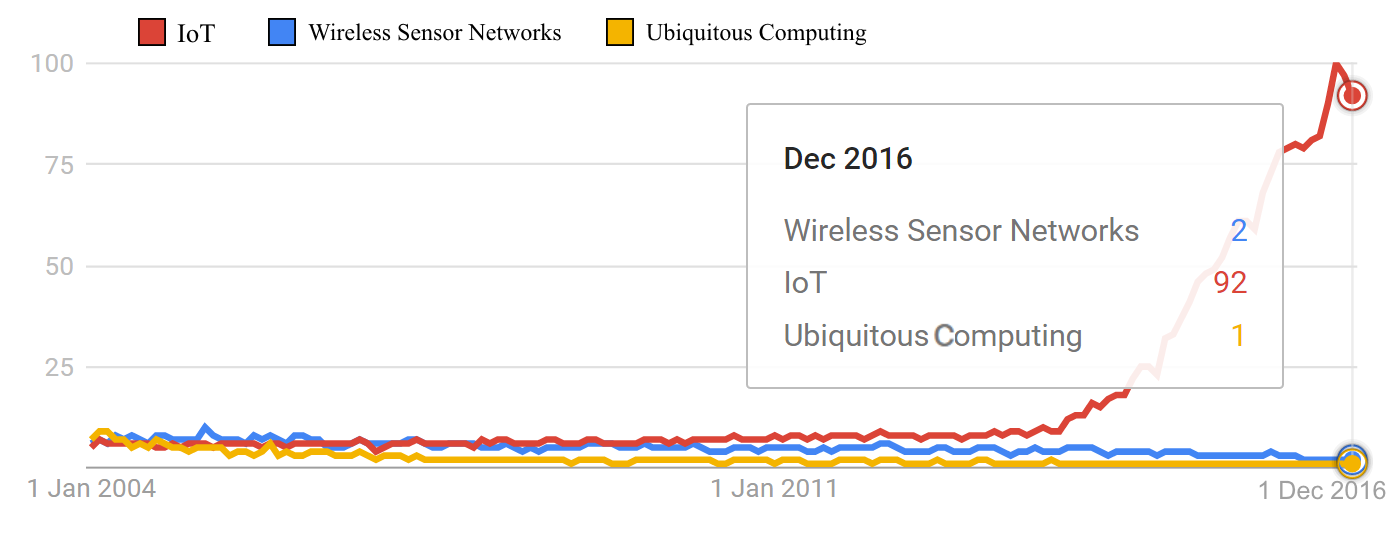}}
	
	\subfloat[Search trends for Wireless Sensor Networks and Ubiquitous Computing]{\includegraphics[width = \textwidth]{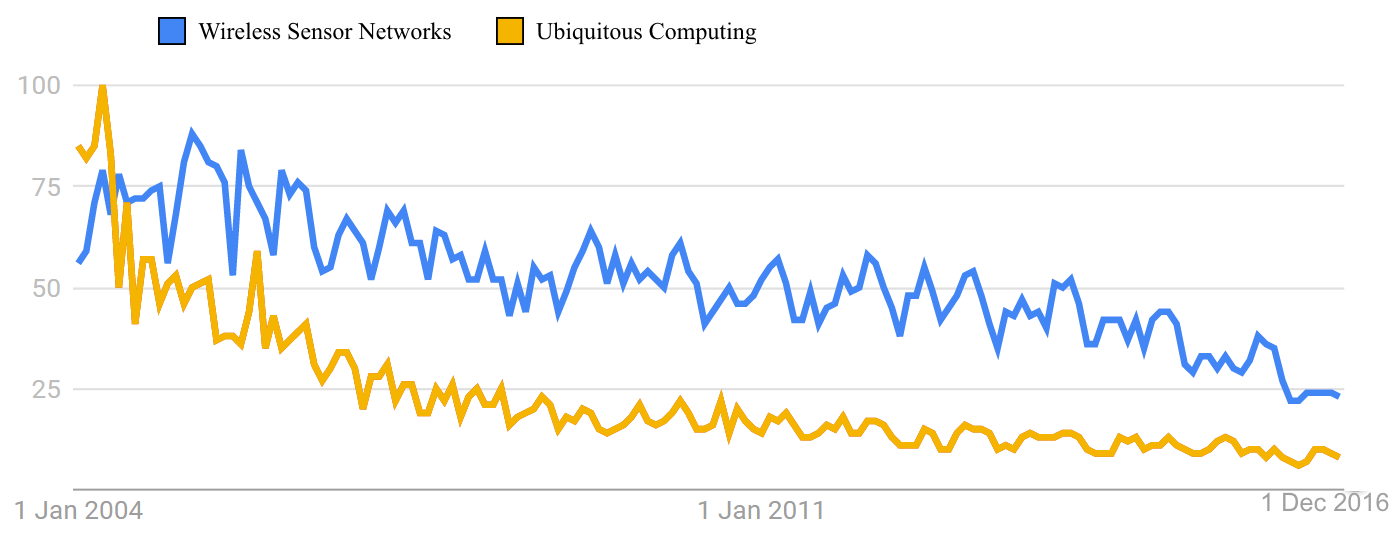}} 
	
	\caption{Google search trends between January 2004 and December 2016: The numbers represent search interest relative to the highest point on the chart for all world and time. A value of 100 is the peak popularity for the term. A value of 50 means that the term is half as popular. Likewise a score of 0 means the term was less than 1\% as popular as the peak \citep{googleTrends}.}
	\label{googleTrends}
\end{figure}

Although different terms have similar meaning the popularity of concepts has changed over the time. Now we look at web search popularity of the terms IoT (including Internet of Things), WSNs and Ubiquitous Computing (UC) as they are used interchangeably. Fig. \ref{googleTrends}(a) is created using Google Trends and it shows how fast the IoT popularity has increased in web searches compared to the terms WSNs and UC over the decade. Similarly, Fig. \ref{googleTrends}(b) shows how the popularity of the terms WSNs and UC has decreased (comparatively) since 2004.

Whether it is called IoT or WSN, there has been a lot of studies to secure these networks starting from the node level to network level \citep{zawoad2015faiot}. The security services provided in IoT and WSN include confidentiality, integrity, authentication, access control, anonymity and availability. 

However, with the increasing prevalence of these devices in many real-life applications, a need has emerged for conducting digital/network forensics to be able to understand the reasons for failures and various attacks. Therefore, in recent years also we have witnessed some studies on cyber-forensics that relate to WSNs or IoT. The goal of this chapter is to investigate such forensic research on WSNs and IoT, and put them in a systematic manner for better understanding and future research. 

This chapter is organized as follows: In the Section \ref{S:df}, we provide a brief background on digital forensics. Section \ref{background} presents related background in IoT and WSNs. Section \ref{app_digital_forensics} discusses how digital forensics can be applied to IoT and WSN environments.

\section{Digital Forensics} \label{S:df}

\begin{figure}
	\centering
	\includegraphics[width=0.8\textwidth,height=\textheight,keepaspectratio]{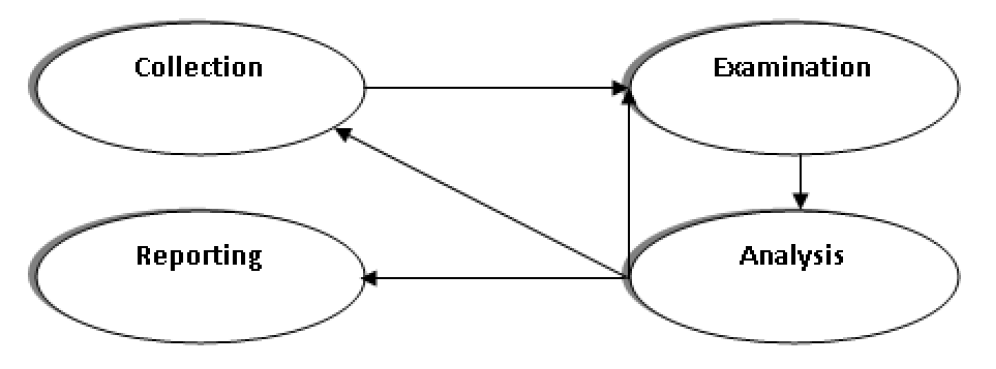}
	\caption{Digital Forensics Process Model \citep{karabiyik2015building}}
	\label{dfprocess}
\end{figure}

Digital Forensics is a branch of forensics science particularly targeting identification, collection (a.k.a. acquisition), examination, analysis, and reporting of digital evidence in order to present it to a court of law. Fig \ref{dfprocess} shows the U.S. Department of Justice's digital forensics investigative process described in ``A guide to first responders" \citep{technical2001electronic}. Digital forensics investigators deal with tremendous amounts of data from numerous types of devices including computers, phones, wearable devices, industrial controls systems, military deployment systems.

When a crime/incident occurs, incident first responders arrive to the scene to identify and secure the digital devices for forensics soundness (preserving integrity of evidence). After securing the evidence devices digital forensics investigators collect digital evidence for further examination and analysis. This basically means to find crime/incident related data on the digital device such as finding traces of an attack and its timestamp on memory of hacked smart TV. During the collection, examination and analysis phases, investigators use digital forensics tools (both hardware and software). These tools help investigators to locate and recover digital evidence which can be both inculpatory (evidence that proves the guilt) and exculpatory (evidence that proves the innocence). At the reporting phase, investigators prepare a report to include in their testimony. When the investigator is asked to testify and present the evidence at a court, the admissibility of the evidence will be questioned based on the procedures followed by the investigator. The most important factor for the admissibility is to verify that the evidence device has not been altered during the investigation. In the case of the IoT environment, this may be quite challenging as there is no universal standard to collect, examine and analyze data from IoT. 

Due to the accelerated advancement in technology, particularly in last the two decades, huge numbers of (heterogeneous) objects became available for personal or enterprise use. This also yields an enormous amount of heterogeneous data and thus more sophisticated and more difficult digital forensics investigations.

\section{Related Background on IoT and WSNs} \label{background}
\begin{figure}[htb]
	\centering
	\includegraphics[width=4.5in,height=0.7in]{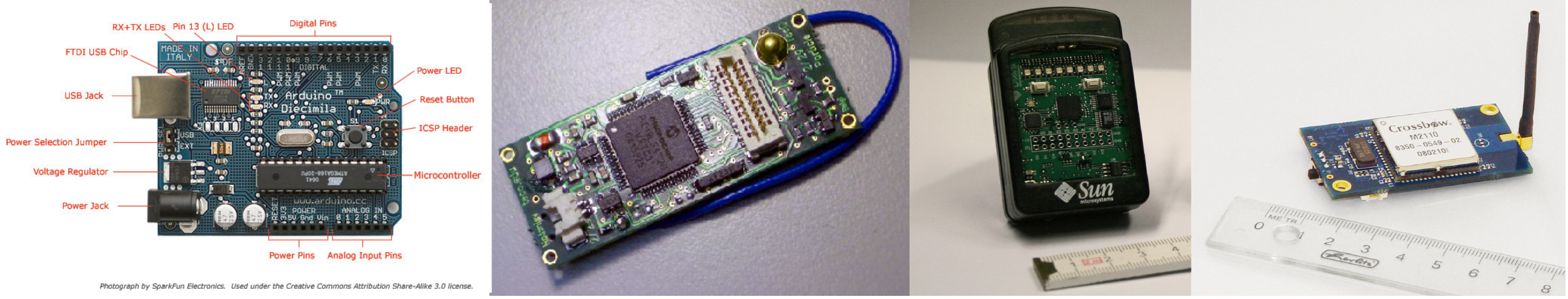}
	\caption{Various sensor platforms: Arduino, Particle, SUN Spots, IRIS Motes}
	\label{sensors}
\end{figure}

The evolutionary background of IoT lies in the advancement of the technology on micro sensor devices in the later 90s. Specifically, the advancements in micro processors, memory technology and more importantly micro sensing devices led to the development of tiny sensors. These sensors are then equipped with radio communication capability on battery energy which enabled unattended intelligent sensing devices that can gather, process and transmit data. In the early 2000s many sensor devices were built to fit the needs of various applications as seen in Fig. \ref{sensors}. 

Of particular interest to these devices are their resources, especially in terms of memory and storage. Early sensors have very scarce resources in terms of memory which makes data storage almost impossible for forensics purposes.  Typically, there were two cases where memory was used: 1) User memory used for storing application-related or personal data; and 2) Program memory used for programming the device. This memory also contains identification data if the device has any. Table \ref{Table:Memory} shows the memory sizes for these devices.

\begin{table}[htb]
\centering
\caption{Memory size for different platforms}
\label{Table:Memory}
\begin{tabular}{ |c|c| }
\hline
 \textbf{Devices} & \textbf{Memory Size} \\
 \hline
  Passive Tags & O(100B) \\
 \hline
  Active Tags & O(1kB)\\
 \hline
  IRIS Motes &  O(100kB)\\
 \hline
  Gateways &  O(100MB) \\
 \hline
\end{tabular}
\end{table}

The other sensor component that is of interest due to forensics would be the communication module. Early devices relied on energy-aware MAC protocols for communication \citep{demirkol2006mac}. Some of them later became standards such as Zigbee \citep{baronti2007wireless} but some of them were only adapted such as Bluetooth \citep{bluetooth2001bluetooth} as seen in Table \ref{Table:Radio}. 

\begin{table}[htb]
\centering
\caption{MAC protocols and data rates for different sensors}
\label{Table:Radio}
\begin{tabular}{ |c|c|c| }
\hline
 \textbf{Standard} & \textbf{Rate} & \textbf{Frequency}  \\
 \hline
  Bluetooth & 2.1Mb/s & 2.4GHz \\
 \hline
  ZigBee & 250kb/s & 2.4GHz/918MHz/868MHz\\
 \hline
  RFM TR 1000 (proprietary) & 19.2kb/s & 916.5MHz\\
 \hline
  Chipcon CC 1000 (proprietary) & 100kb/s & 433MHz \\
 \hline
\end{tabular}
\end{table}

At the network layer, energy-efficient routing protocols were developed to provide service to large-scale WSNs consisting of thousands of nodes \citep{werner2006deploying,szewczyk2004analysis,arora2004line} and employ multi-hop communication \citep{akkaya2005survey}. Zigbee Alliance had also routing and application layer protocols for WSNs \citep{baronti2007wireless}. In most cases, these protocols were distributed and required sensor nodes to maintain a simple routing table for data forwarding. In some cases, the routing protocol was managed by the gateway which is assumed to collect all the sensor data from the sensors. In any case, there was enough information in the sensors or gateway to be able to identify routing failures in real-time but this might be challenging for cyberforensics purposes as will be discussed later.  

The heavy research on WSNs later led to the development of some standards such as Zigbee/IEEE802.15.4, IETF ROLL \citep{ko2011connecting,sheng2013survey}, IETF 6LoWPAN \citep{shelby20116lowpan}, Wireless HART \citep{song2008wirelesshart}, ISA100 \citep{isa1002009isa100} which accelerated the production of sensor devices. In the meantime, there has been further developments to enrich the resources of sensor devices and getting them connected to the Internet. The enrichment was in terms of processor and memory capacity and the number of sensing modules. With the proliferation of smart mobile phones, the idea of smart, connected, sensing and battery-operated devices have penetrated our everyday lives which has led to the development of similar products to make our lives convenient. Within a few years, a lot of sensing and communication capable devices such as smart meters, cameras, thermostats, wearables, RFIDs, tags, bulbs, beds, speakers, locks, watches, cookers, keypad, and applicances, have started to be seen which are referred to as IoT devices in general \citep{iot}. 

With the enriched resources, these devices started to look like more of our laptops with comparable memory/storage sizes and communication capabilities. In addition to Zigbee or Bluetooth, Wifi/4G has also started to be used for communication purposes. Finally, the data collected from these devices was not stored in the gateways but rather transferred to cloud storage where it can be accessed for later use. 

The IoT era changed the needs of the WSN era and Digital Forensics was one of the affected area as the devices are being used in a lot of daily applications by humans. Therefore, we discuss how digital forensics is applied to the IoT and WSNs.

\section{Applying Digital Forensics to IoT and WSNs} \label{app_digital_forensics}

IoT and WSNs consist of sensitive data stored and processed hence, in theory it is suggested that the data which is processed and cumulated by well known firms will be the subject of future digital forensic investigations. The evidence that is provided by IoT or WSNs to the forensic community will be far more finer compared to what the community currently possesses. In addition, IoT and WSNs also offer new and better opportunities for data that is at times misused, through growth and development in the forensic community's procedures. The techniques/algorithms methods that were used and or developed were based on the digital forensics process model consisting of collection, examination, analysis, and reporting of the data/evidence. Using these practices not only data for evidence is identified in a myriad system, but is also preserved for future references as the information presented is an intense fusion of collection, extraction, processing, and interpretations.

Digital forensics in IoT/WSNs is a challenge especially when it comes to accuracy due to the intensity of analysis. This results in data sometimes losing its granularity as systems may store, use, or present different semantics however, it does have the ability to adopt dissimilar formats, and may hold a proprietary format. Taking into the heterogeneity of data that IoT/WSNs devices generate it is even more challenging. The following questions must be answered before the investigation is being performed in order to avoid inadmissibility of evidence. Can data be collected from the devices using available tools? Is the data propriety? How can it be analyzed? Are forensic tools compatible with this data?   

Most of the challenges in IoT forensics are also available to the WSNs particularly at the device/data storage and network levels. The only difference in most cases is the scale of WSNs because of the application-specific needs. Early WSNs works lacked any security in regards to integrity and authentication because of the broadcast nature of communication. There was no formal set of requirements for achieving forensic readiness in WSNs. However, with the rapid tendency towards the usage of efficient, low memory  footprint and low power devices in the industry, devices will be less likely to keep data stored in memory. Therefore, similar forensic readiness frameworks that will be discussed in the following sections must be developed for such devices in advance. Otherwise, forensically crucial data can be easily lost forever.   

In the next section we will discuss the challenges that investigators and practitioners face when performing digital forensics procedures in both IoT and WSNs. Although IoT and WSNs are different with respect to their structures, WSNs are considered to be part of IoT \citep{christin2009wireless,mainetti2011evolution,li2013practical,IOTandWSN}, a concept of worldwide connected ubiquitous devices. The distinctions between two environments are not clearly pointed by the research in the current literature and digital forensics efforts similarly applied to both concepts, particularly research in IoT forensics are conducted with WSN characteristics in mind. This makes some of the research efforts in both environments inseparable from the digital forensics perspectives.

\subsection{Challenges in IoT and WSN Forensics} \label{challenges}

In this section, we discuss the digital forensics challenges for IoT and WSN as specified by \citet{hegarty2014digital}. Note that most of the challenges we discuss in this section are applicable to both IoT and WSN.

\begin{table}[htb]
\centering
\caption{Potential evidence sources in IoT and WSN Environments \citep{oriwoh2013forensics}}
\label{tab:iotEvidence}
\setlength\extrarowheight{2pt}
\begin{tabular}{|c|l|l|l|}
\hline
\multicolumn{1}{|l|}{}                                                                   & \multicolumn{1}{c|}{\textbf{Sources}}                        & \multicolumn{1}{c|}{\textbf{Example}}                                                                                                                                & \multicolumn{1}{c|}{\textbf{\begin{tabular}[c]{@{}c@{}}Expected\\ evidence\end{tabular}}}              \\ \hline
\multirow{3}{*}{\textbf{\begin{tabular}[c]{@{}c@{}}Internal to \\ network\end{tabular}}} & \begin{tabular}[c]{@{}l@{}}Hardware End\\ nodes\end{tabular} & \begin{tabular}[c]{@{}l@{}}IoTware e.g. game consoles, fridges, \\ mobile devices, smart meters, \\ readers, tags, embedded systems, \\ heat controller\end{tabular} & \begin{tabular}[c]{@{}l@{}}Sensor data e.g.\\ IP address, Rime \\ number, sensor ID\end{tabular} \\ \cline{2-4} 
                                                                                         & Network                                                      & \begin{tabular}[c]{@{}l@{}}Wired and Wireless, mobile \\ communications e.g. GSM, sensor\\ networks, HIDS, NIDS, HMS\end{tabular}                                    & Network, Logs                                                                                          \\ \cline{2-4} 
                                                                                         & \begin{tabular}[c]{@{}l@{}}Perimeter \\ devices\end{tabular} & \begin{tabular}[c]{@{}l@{}}AAA server, firewall, NAT server, \\ IDS, NIDS, HIDS\end{tabular}                                                                         &                                                                                                        \\ \hline
\multirow{4}{*}{\textbf{External}}                                                       & Cloud                                                        & \begin{tabular}[c]{@{}l@{}}Public, Private, Hybrid cloud \\ systems\end{tabular}                                                                                     & \begin{tabular}[c]{@{}l@{}}Client Virtual \\ Machines; logs\end{tabular}                               \\ \cline{2-4} 
                                                                                         & Web                                                          & \begin{tabular}[c]{@{}l@{}}Web clients, webservers, \\ social networks\end{tabular}                                                                                  & \begin{tabular}[c]{@{}l@{}}Web logs; user \\ activity\end{tabular}                                     \\ \cline{2-4} 
                                                                                         & \begin{tabular}[c]{@{}l@{}}Hardware End\\ nodes\end{tabular} & \begin{tabular}[c]{@{}l@{}}Mobile devices, sensor nodes \\ and networks\end{tabular}                                                                                 & \begin{tabular}[c]{@{}l@{}}Sensor data e.g.\\ IP address, Rime \\ number, sensor ID\end{tabular} \\ \cline{2-4} 
                                                                                         & \begin{tabular}[c]{@{}l@{}}`X' Area\\ Networks\end{tabular}  & Home Area Networks (HAN)                                                                                                                                             & Network logs                                                                                           \\ \hline
\end{tabular}
\end{table}

\subsubsection{Different Interfaces and Storage Units:} The IoT devices that are used in everyday life have different interfaces which allow users to use services or control the devices. Example interfaces could be propriety software, mobile application, hardware, or embedded firmware which provides an invisible interface. The variety of interfaces makes digital forensic investigation a tedious process as digital forensics tools do not automatically detect all types of interfaces, file systems and even data itself. Similar issues arise when WSNs are the forensically targeted environments. In addition to the variety of interfaces, IoT devices store data in miscellany of storage units both volatile and non-volatile including internal and external memory units (e.g. eMMC, eFlash, and DRAM) and cloud storage (e.g. HDD and SSD) \citep{pereira2013enabling}. As for the sources of digital evidence, Table \ref{tab:iotEvidence} gives a broader view of where potential evidence may reside in an IoT and WSN environment.   

Differences in the interfaces and storage units causes investigators to perform manual forensic methods on the devices if (at all) possible. This will also increase the time required for the investigation as automated tools do not recognize propriety interfaces. Another issue is that volatile data might be destroyed by the device after they are used. In this case, data recovery may not be even possible. In addition, data may also be destroyed due to the wear-leveling technology in flash memory devices and solid state drives. Every memory cell has a certain read/write lifetime which varies between 10000 and 100000 depending on the manufacturer. Therefore, internal firmware in the memory will distribute data stored in the memory to the unused (unallocated) cells in order to level memory wearing in mostly used memory cells. In this case, previously deleted data will be destroyed because unallocated space also consists of the memory cells that has previously been used to store data but later deleted.

\begin{figure}[htb]
	\centering
	\includegraphics[width=0.6\textwidth,height=\textheight,keepaspectratio]{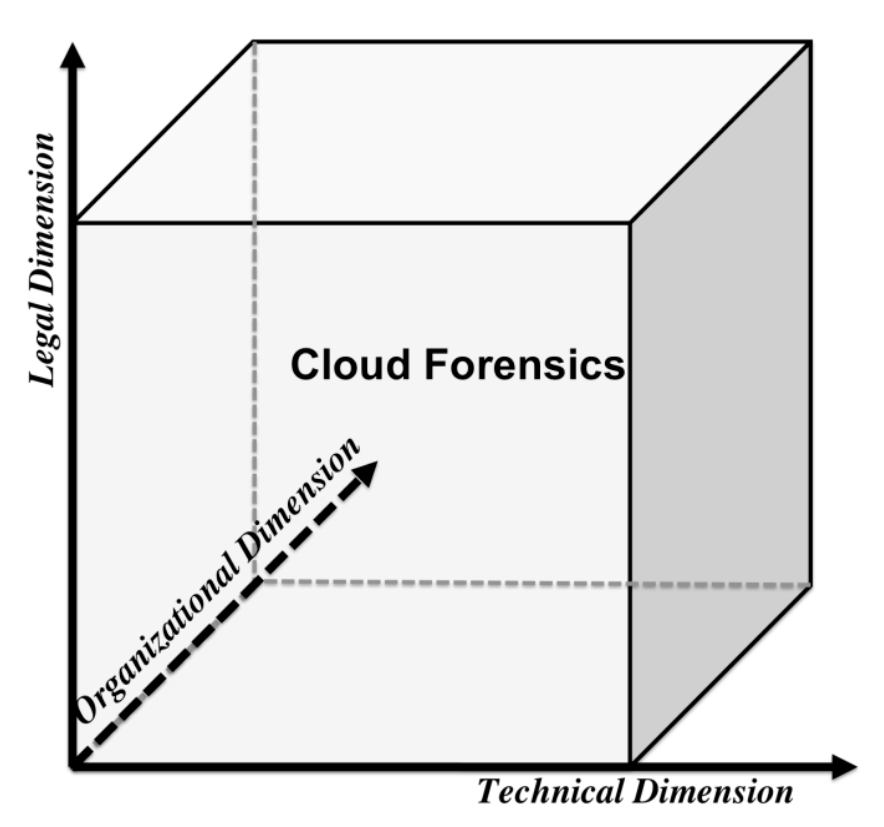}
	\caption{Cloud forensic three-dimensional model \citep{ruan2011cloud}}
	\label{cloudForensics}
\end{figure}

Furthermore, data stored in the cloud raises serious issues in digital forensics investigations performed in IoT and WSN environment. In order to identify these issues, \citet{ruan2011cloud} defines cloud forensics in three dimensions (see Fig. \ref{cloudForensics}). The two most important problems in cloud forensics that are also highly related to IoT and WSN forensics are multi-tenancy and multi-jurisdiction. Multi-tenancy allows cloud tenants to access the software instance simultaneously \citep{computing2010toward}, therefore user ascription and ownership for specific data become the investigators' major concerns.

In order to provide efficient service availability and reduce the cost of services, major cloud service providers such as Google, Amazon and HP locate their data centers all around the world. Different countries and different states have different jurisdictions. A crime in one jurisdiction may not be considered a crime in another. In addition, law enforcement agencies having different jurisdictions may not be willing to cooperate with each other. Due to all of these issues, investigators may have to deal with multi-jurisdiction issues when data from IoT and WSNs are stored in cloud.

\subsubsection{Lack of Universal Standard for IoT and WSN Data Storage}
Due to the characteristics of IoT/WSN data, it is extremely difficult to create a universal standard for data storage. Nevertheless, there have been efforts to create frameworks to provide a unified way to store data for IoT. \citet{li2012storage} identify the IoT data features as follows:

\begin{itemize}
	\item \textbf{\textit{Multi-source and Heterogeneity:}} IoT/WSN data is sampled by various connected devices including Radio Frequency Identification (RFID) readers, cameras, smart appliances, proximity, pressure, temperature, humidity, and smoke sensors. The data collected from this vast category of devices have significantly different semantics and structures. 
	\item \textbf{\textit{Huge scale:}}  The IoT/WSN contains a large number of perception devices, these devices' continuously and	automatically collect information leads to a rapid expansion of data scale.
	\item \textbf{\textit{Temporal-spatial correlation:}} As the data are constantly collected from IoT/WSN, the data will consist both time and space attributes in order to correlate them with respect to the changing location of device over time.
	\item \textbf{\textit{Interoperability:}} IoT/WSN are currently evolving to achieve data-sharing to facilitate collaborative work between different applications. For instance, in the case of an on-road emergency, while the patient's medical record is securely shared with a nearest emergency center \citep{rabieh2018secure}, the data related to road conditions may be also assessed for timely arrival by an autonomous car.
	\item \textbf{\textit{Multi-dimensional:}} IoT application now integrates several sensors or WNSs to simultaneously monitor a number of sensing devices, such as temperature, humidity, light, pressure, and so on. And thus the sample data is usually multidimensional.
\end{itemize}

The available methods and techniques are mostly limited and designed for a certain set of technologies. For instance \citet{li2012storage} have proposed a solution to the storage and  management of IoT data named IOTMDB using NoSQL (Not Only SQL). In addition to this work, \citet{jiang2014iot} proposed a data storage framework to efficiently store big IoT data which is collected from the deployed devices (WSNs) into storage units by combining and extending multiple databases and \textit{Hadoop} (an open source framework that provides capability of process and storage of large data sets). In addition, \citet{gubbi2013internet} introduced a conceptual IoT framework with \textit{Aneka} cloud computing platform-runtime platform and a framework for developing distributed applications on the cloud \citep{aneka}- being at the center. This framework integrates ubiquitous sensors and various applications (e.g. surveillance, health monitoring and critical infrastructure monitoring) using aforementioned cloud platform.  

From the forensics investigation's point of view, analysis of data coming from different sources will be a serious challenge. The only way to deal with the analysis of such heterogeneous data is to use Hexadecimal editors (a.k.a HEX editors) as they allow reading the raw data from storage units. However, it will be a tedious (if not infeasible with large scale data) process because of the amount of data collected from IoT devices and WSNs. 

Temporal-spatial correlation of IoT/WSN data may be useful for the investigators when data includes geolocation information (e.g., GPS coordinates) readable by the tools used. However, IoT/WSN space can be defined of any size and data may come with custom space information. This also needs to be translated into intelligible data by the investigators as evidence. 

Interoperability of the devices will be a serious challenge for forensics investigations as the data will be shared among the applications and the origin of the data needs to be known to conclude the investigation. If the data being operated by different applications is not traceable then accountability or non-repudiation issues will be raised. For instance, it will be difficult to answer the questions: What caused the operation failure? Was there any attack? What data is produced by each application/device?

\subsubsection{Devices have different levels of complexity, battery life/source}

As discussed earlier, IoT devices may vary depending on the duties they perform, how often the device communicates, size of the data being transmitted, and available storage in the device \citep{siliconlabs}. This variance is also reflected on the complexity of devices. While the device may be as simple as a single sensor collecting environmental values from animals' habitats, it may also be complex enough to consist of a processor, relatively large memory units, and communication protocols with security mechanisms (e.g., Internet refrigerator). In the former, battery replacement will be impractical therefore battery life is expected to outlive the animal \citep{chen2012challenges}. In the latter however, the device will need to constantly consume power to be available for its service. 

Complexity and battery life/source of the device affects digital forensics investigations from similar points of view as discussed above such as volatility of data, availability of data, ownership and user ascription. For example, the data in network and volatile memory disappears in a short amount of time, thus recovery of such data is often impossible unless the device keeps logs of data. This requires existence of more non-volatile memory and processing power hence larger battery.

\begin{figure}[htb]
	\centering
	\includegraphics[width=0.75\textwidth,height=\textheight,keepaspectratio]{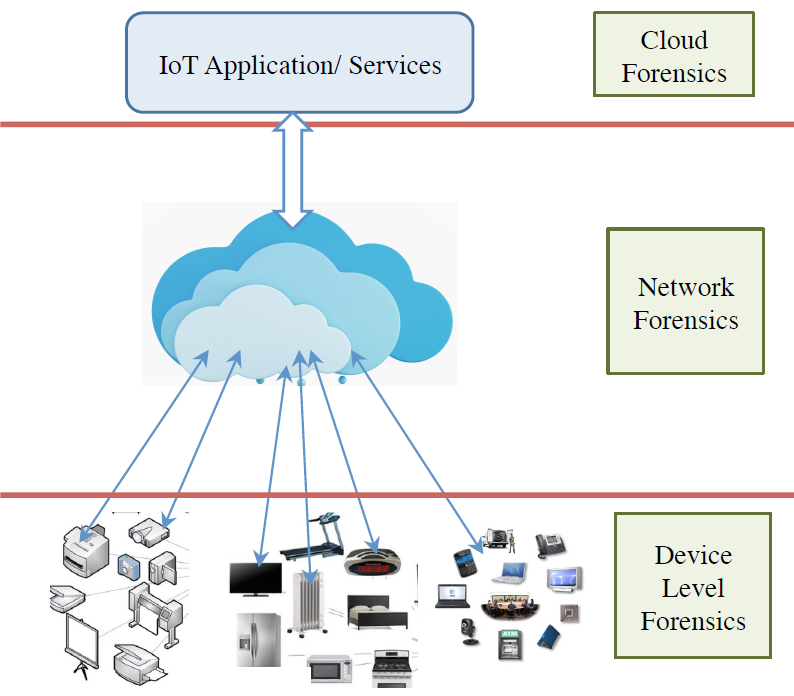}
    \caption{IoT Forensics \citep{zawoad2015faiot}}
    \label{iotForensics}
\end{figure}

\subsubsection{Availability of Propriety Operating Systems} 

The operating systems (OS) that are used for IoT was originally designed for WSNs such as TinyOS \citep{levis2005tinyos}, Contiki \citep{dunkels2004contiki} and OpenEmbedded Linux \citep{openembedded}. However, with the advances in the development of more sophisticated IoT devices than small sensors, the need for new OSs for IoT emerged. Hence, RIOT \citep{baccelli2013riot} was developed to bridge the gap between the available OSs for WSNs and the new needs for IoT. Recent development of Android Things \citep{androidThings} also move this trend to another level to provide leveraging the existing Android development tools, APIs and resources to build an IoT environment. While Table \ref{tab:oss} depicts the comparison of different OSs for IoT and WSN, the details about these existing solutions can be obtained from the given resources above. In Table \ref{tab:oss}, P means: Partial Support, N means: No Support and Y means: Full Support for given points. 

\begin{table}[htb]
\centering
\caption{Comparison tables for different IoT and WSN Operating Systems and supported protocols \citep{minerva2015towards}}
\label{tab:oss}
\setlength\extrarowheight{3pt}
\begin{tabular}{l|ccccccc}
\hline
\textbf{OS} & \textbf{\begin{tabular}[c]{@{}c@{}}Min\\ RAM\end{tabular}} & \textbf{\begin{tabular}[c]{@{}c@{}}Min \\ ROM\end{tabular}} & \textbf{\begin{tabular}[c]{@{}c@{}}C \\ Support\end{tabular}} & \textbf{\begin{tabular}[c]{@{}c@{}}C++ \\ Support\end{tabular}} & \textbf{\begin{tabular}[c]{@{}c@{}}Multi-\\ Threading\end{tabular}} & \textbf{Modularity} & \textbf{Real-Time} \\ \hline
\textbf{Contiki} & \textless2KB & \textless30KB & P & N & P & P & P \\ 
\textbf{TinyOS} & \textless1kB & \textless4kB & N & N & P & N & N \\ 
\textbf{RIOT} & $\sim$1.5kB & $\sim$5kB & Y & Y & Y & Y & Y \\ 
\textbf{Linux} & $\sim$1MB & $\sim$1MB & Y & Y & Y & Y & P \\ \hline
\end{tabular}

\bigskip 

\begin{tabular}{l|ccccc}
\hline
\textbf{OS}      & \textbf{IPv6} & \textbf{TCP} & \textbf{6LoWPAN} & \textbf{RPL} & \textbf{CoAP} \\ \hline
\textbf{Contiki} & Y             & P            & Y                & Y            & Y             \\
\textbf{TinyOS} & N             & P            & Y                & Y            & Y             \\
\textbf{RIOT}    & Y             & Y            & Y                & P            & P             \\
\textbf{Linux}   & Y             & Y            & Y                & Y            & N            \\ \hline
\end{tabular}

\end{table}

There has been digital forensics research on the protocols such as IPv6 \citep{nikkel2007introduction,kumar2014digital,kumar2016traffic}, 6LoWPAN \citep{perumal2015internet, kumar2016traffic}, and RPL \citep{kumar2016traffic} that we mentioned in Table \ref{tab:oss}. These research efforts mostly provide frameworks for forensic readiness of IoT and WSNs. To the contrary of the availability of forensic readiness frameworks, wide variety of available protocols for both IoT and WSN creates a troublesome investigative process and introduces a steep learning curve for forensic examiners.

IoT forensics can be divided into three categories depending on where the forensic data is located and the investigation can take place (see Fig. \ref{iotForensics}) \citep{zawoad2015faiot}. Specifically these are : 1) Device/Node; 2) Network where the data is collected; and 3) Cloud where the data is stored. The forensic research on WSNs is conducted at the first and second levels where sensor data is collected and transfered, and the communication takes place. Next we explain each category.

\subsection{Device Level Investigation} \label{devicelevel}

IoT or WSNs deploy a variety of devices with certain characteristics. Typically these devices employ processing units, memory, communication module and sensing modules. The richness of the set of such devices increased significantly with the developments in micro electro-mechanical devices \citep{gaura2006smart}. Examples of devices include but is not limited to sensors, smart phones, smart meters, smart thermostats, cameras, wearable devices, on-board vehicle devices, RFIDs, smart watches, and drones. 

Device level investigation is necessary when data needs to be collected from the memory of a device in IoT/WSN. As discussed in Section \ref{challenges}, IoT/WSN devices may have propriety interfaces and storage units. Although it creates a burden on investigators in terms of longer investigation time and increased learning curve, evidence must be collected from these heterogeneous devices. Thus, the current state of the research shows that there is a long way to standardize the device level investigations for both IoT and WSNs environments. In this section, we explain general forensics techniques which are used when data is not available through device's interface. We then discuss some of the techniques used to collect forensic data from specific devices and their memories.

The National Institute of Standards and Technology (NIST) discussed different digital forensics data acquisition techniques from mobile devices in ``Guidelines on Mobile Device Forensics" \citep{ayers2013guidelines}. They recommend performing the following acquisition methods: manual extraction, logical extraction, hex dumping/JTAG, chip-off, micro read. Manual and logical acquisition methods are usually available when devices provide user interface and are not locked, password protected, and damaged. In the case of IoT devices (other than smartphones and tablets), it is mostly not the case. Therefore, investigators usually perform hex dumping/JTAG and chip-off techniques (micro read is a special technique and it is out of our scope). 

When smartphones or tablets are the interests of the investigations, examiners use state-of-art digital forensic tools such as Cellebrite UFED Physical Analyzer, Paraben Device Seizure, XRY, and Oxygen Forensics for their data acquisition and analysis. This is mainly because these devices come with a well understood operating systems such as Android, iOS, or Windows. Therefore, physical and logical acquisition is generally available to the investigators using aforementioned toolkits. Although we discuss some data acquisition techniques from mobile devices in this section, we do not elaborate more on the available forensic toolkits.  

Forensically related data from a mobile device's main storage unit is typically available for acquisition, however volatile data acquisition could often be challenging. Therefore, particular research interest from the mobile forensics community emerged for volatile memory acquisition. \cite{anderson2008white}, \cite{kollar2010forensic}, and \cite{sylve2012acquisition} proposed early forensic volatile memory dumping tools \textit{crash}, \textit{fmem}, and \textit{dmd} respectively. The acquired data from these tools is then analyzed using other available tools such as hex editors. As a more recent research, \cite{saltaformaggio2016screen} proposed an open source tool called RetroScope which recovers multiple previous screens (from 3 to 11) from the volatile memory of a smartphone using  a spatial-temporal memory acquisition technique. This technique shows that investigators can recover earlier content of an application (e.g. Facebook, WhatsApp, and WeChat) after the data is not available through conventional techniques and tools. This technique can also be particularly effective when the investigators do not have access to the smartphone's data storage due to being password protected. Another recent research on volatile memory acquisition tool development for mobile devices is done by \cite{yang2016tool}. The proposed tool, AMExtractor, collects volatile memory from a wide variety of Android devices for forensic acquisition meaning with high integrity.   

\begin{figure}[htb]
	\centering
	\includegraphics[width=0.75\textwidth,height=\textheight,keepaspectratio]{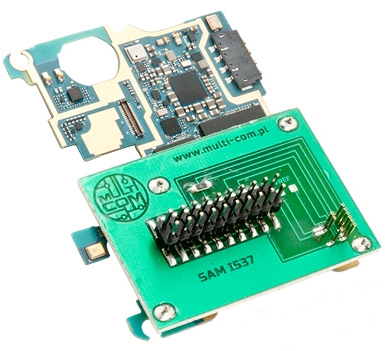}
	\caption{JTAG module for Samsung Galaxy S4 Active smartphone}
	\label{jtag}
\end{figure}

Hex dumping/JTAG technique allows investigators to access the memory content by connecting special cables to the provided pins on the device. This is done by loading a firmware to the device's memory which is then used to access the information in the rest of the device memory. Fig. \ref{jtag} shows the JTAG module attached to Samsung Galaxy S4 Active phone's mainboard. Using the connectors available on the module and forensic memory reading tools, data from the phone's memory can be easily accessed.

Chip-off is another technique used when phone data is not available due to several reasons such as JTAG is not possible and phones being physically broken, burned or locked. In such cases,  investigators can physically remove the flash memory from the device using chip-off technique. Although this techniques is described for mobile phones, it can also be used pretty much for any IoT device or a sensor in WSNs which stores data in flash memory (NAND, NOR, OneNAND or eMMC) \citep{binaryintel}. It is also important to note that chip-off techniques may damage the memory and may cause permanent data loss even though all the precautions are taken \citep{swauger2012chip}.

\begin{figure}[htb]
	\subfloat[iPhone 5c with removed NAND] {\includegraphics[width=0.49\textwidth,height=\textheight,keepaspectratio]{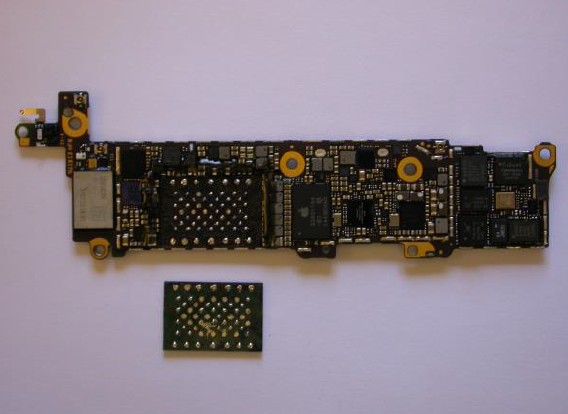}}
	~~~
	\subfloat[Test board for copying NAND chips] {\includegraphics[width=0.48\textwidth,height=\textheight,keepaspectratio]{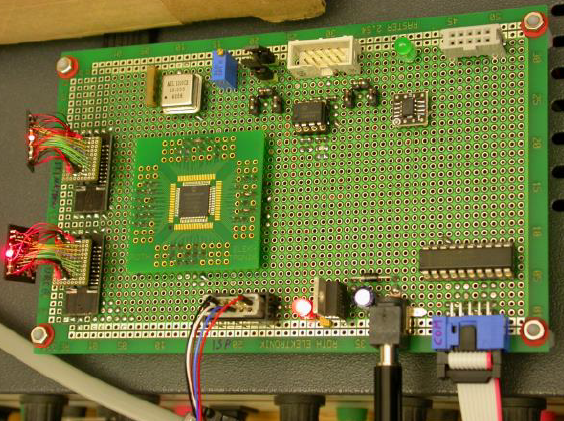}} 
	
	\caption{Chip-off and NAND memory mirroring for iPhone 5c (both figures are from \citep{skorobogatov2016bumpy})}
	\label{fig:iphone}
\end{figure}

\begin{figure}[htb]
	\centering
	\includegraphics[width=0.8\textwidth,height=\textheight,keepaspectratio]{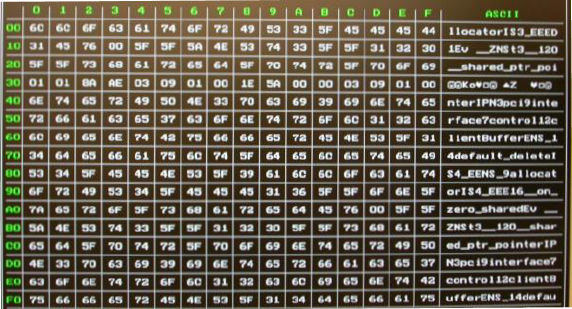}
	\caption{Data acquired from iPhone 5c NAND via reading software \citep{skorobogatov2016bumpy}}
	\label{fig:nanddata}
\end{figure}

Chip-off is a delicate and challenging method of data acquisition, therefore it requires extensive training in both electronic engineering and file system forensics. After the memory is removed from the phone, investigators are able to create binary image (bit-by-bit copy) of the removed memory. Fig. \ref{fig:iphone}(a) shows removed NAND flash memory from iPhone5c and Fig. \ref{fig:iphone}(b) shows example of how removed NAND chip is mirrored using a test board. Finally, Fig. \ref{fig:nanddata} shows the raw data acquired from the removed memory via chip programmer and reading program. 

\citet{zaharis2010live} propose an architecture which provides remote live forensics protection and eliminates malicious code execution in WSNs using sandboxing methods. Using the proposed architecture, one may dump the volatile memory from the sensor device. However, this architecture does not provide full memory dump for analysis, instead it extracts data selectively due to power efficiency constraints. The collected data is only used for verification of the integrity of the program that each sensor device is running. Nevertheless, this is not considered complete forensics analysis of sensor device memory.

In order to close the gap discussed above, \citet{kumar2014digital} propose an architecture of memory extraction from devices that are used in both IoT and WSNs environments. The main goal of this work is to investigate the extracted data in order to determine the reasons which could have caused the security breaches. This architecture is specifically designed to extract, analyze, and correlate forensic data for IPv6-based WNS devices which run Contiki \citep{dunkels2004contiki} operating system and is powered by 8051-based, 8-bit micro-controllers. Contiki is a light-weight and open source operating system for IoT and WSN devices. It is important to note that the analysis done by the authors is purely hardware based and does not depend on WSN traffic analysis. 

This work is divided into three steps which are extraction, analysis, and co-relation. In the first step, a copy of memory is extracted from the device memory. In the second step, the acquired data is analyzed in a fully automated fashion in order to reduce investigation time. In the final step, a set of new data is looked for by co-relating retrieved data from one device to another device in the case of multiple devices being used in the network.

As wearable IoT devices are becoming part of our everyday life, especially fitness trackers, they started to appear in the crime/incident scene and also being used in court cases \citep{fitnessNews1, fitnessNews2, fitnessNews3}. This resulted in the need for forensic data collection from fitness trackers with different interfaces. \citet{cyr2014security} have studied security analysis of Fitbit, a wearable fitness device. Although they mostly focused on the security issues in both device communication and mobile application, its importance is also negligible from the digital forensics perspective. This is mainly because their methods can be used by forensic investigators. 
  
\begin{figure}[htb]
	\centering
	\includegraphics[width=\textwidth,height=\textheight,keepaspectratio]{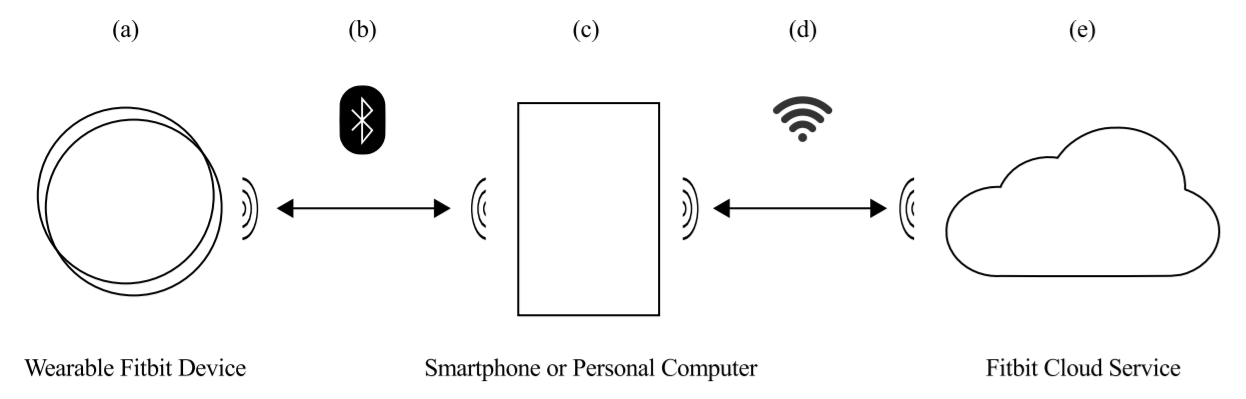}
	\caption{The Fitbit system components: attack surface is partitioned into five regions, (a), (b), (c), (d), and (e) \citep{cyr2014security}}
	\label{fig:fitbitsystem}
\end{figure}

Fig.~\ref{fig:fitbitsystem} shows each component in Fitbit system when synchronization is performed between a Fitbit device, mobile device or computer, and Fitbit cloud service. This system is also partitioned into possible attack surfaces in the figure. In addition to security analysis, the same partitioning can also be used for forensic investigation as well. The device's memory can be extracted from Fig.~\ref{fig:fitbitsystem}(a) and analyzed using JTAG or chip-off techniques, and a chip reading software. Fig.~\ref{fig:fitbitsystem}(b) and Fig.~\ref{fig:fitbitsystem}(c) can be attacked to read communication between both devices shown in Fig.~\ref{fig:fitbitsystem}(a) and Fig.~\ref{fig:fitbitsystem}(c). Fitbit cloud data however, can be retrieved using similar methods discussed later in Section~\ref{cloudLevel}.

\begin{figure}[htb]
	\includegraphics[width=0.48\textwidth,height=\textheight,keepaspectratio]{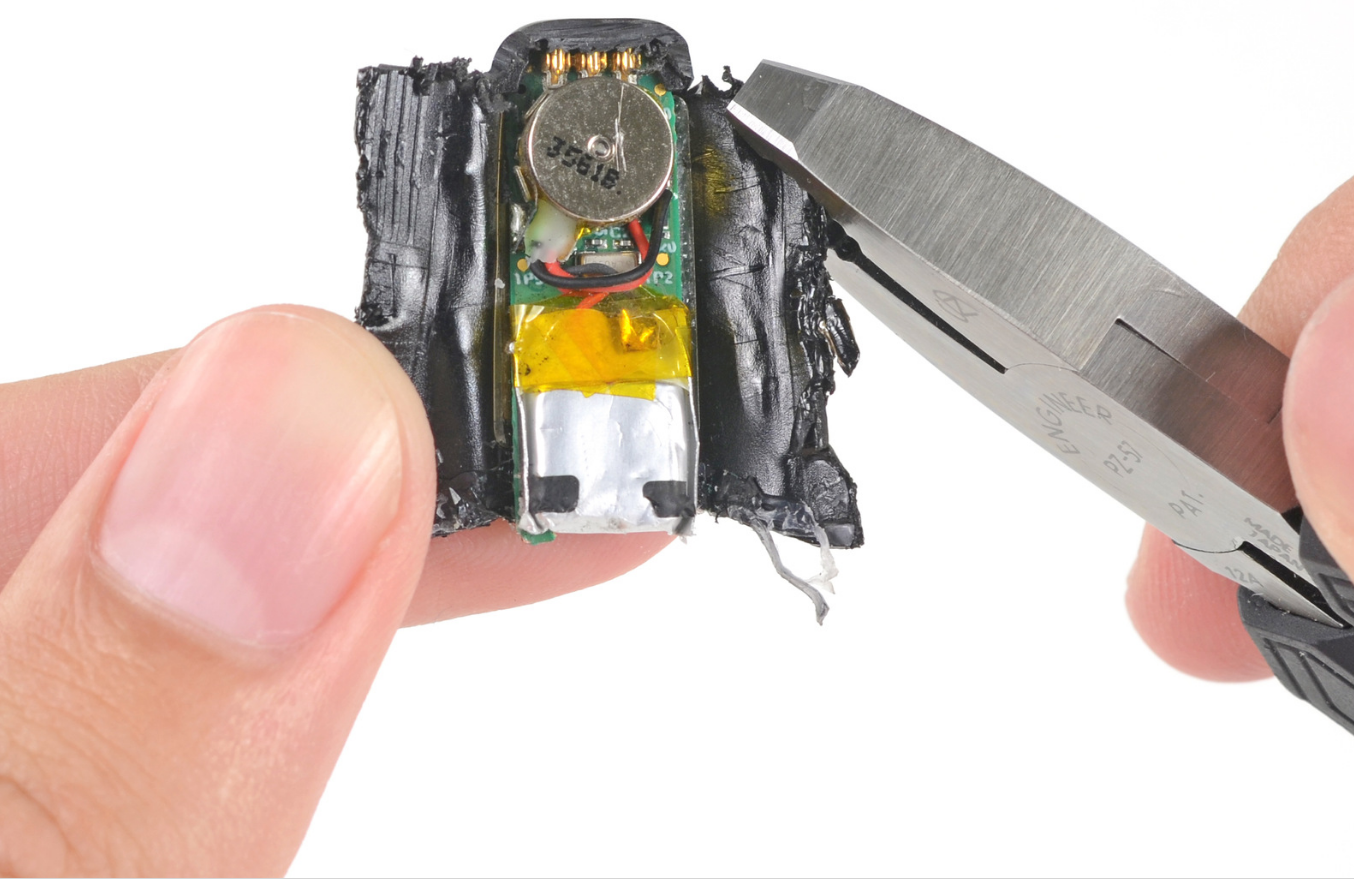}
	~~~
	\includegraphics[width=0.49\textwidth,height=\textheight,keepaspectratio]{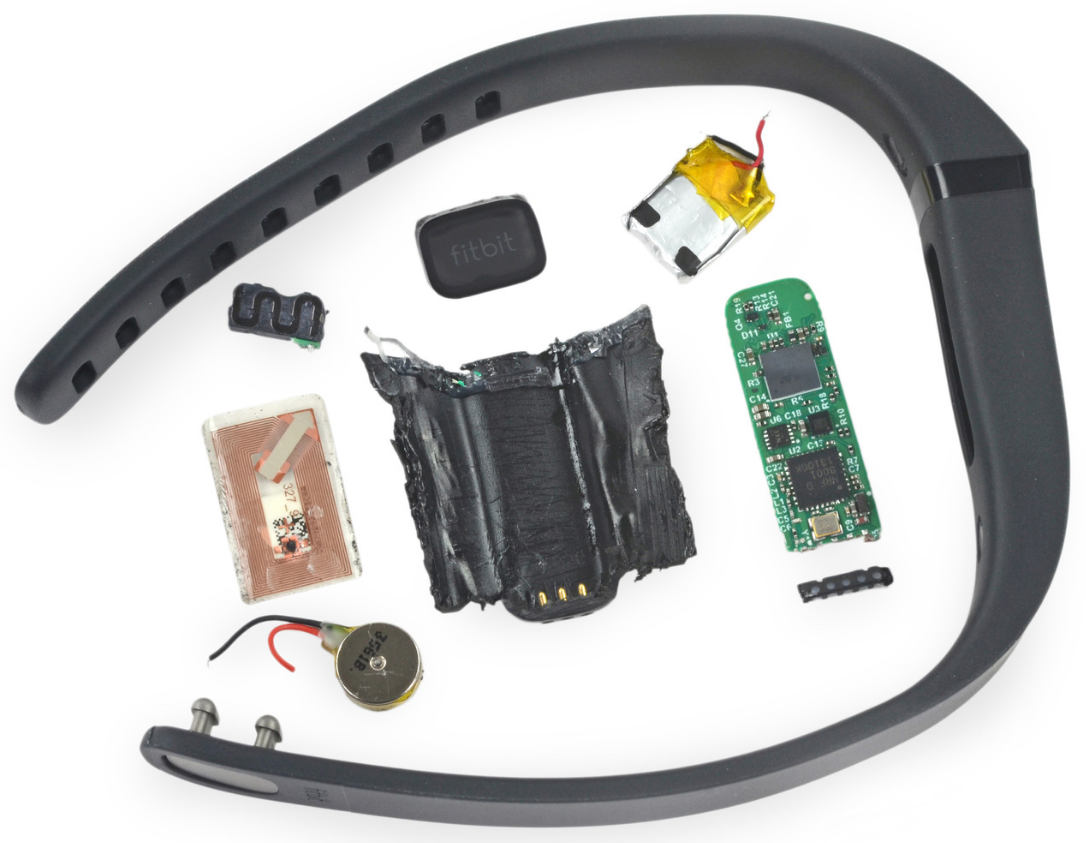}
	\caption{Fitbit Flex teardown process \citep{ifixit}}
	\label{fitbitteardown}
\end{figure}

In most of the wearable fitness devices, memory is packaged with waterproof material. Therefore, it is impossible to physically access the memory without destroying the packaging (see Fig. \ref{fitbitteardown}). Once the memory device is accessed, then JTAG or chip-off can be used to retrieve raw data from the memory.

\subsection{Network Level Investigation} \label{networklevel}

In some applications, IoT devices or sensors form a network of collective sensing and action. Therefore, in addition to device level data, there will be data collected at the network level regarding the flow of data, routing and tracking of lost packets. This IoT-related network may utilize one or more of the following networks:
\begin{itemize}
	\item Body Area Network (BAN), 
	\item Personal Area	Network (PAN),
	\item Home/Hospital Area Networks (HAN),
	\item Local Area Networks (LAN),
	\item Wide Area Networks (WAN),
    \item Cyber-Physical System (CPS).
\end{itemize}

For each type of network, there needs to be customized mechanisms to be able to conduct cyber forensics after an incident. This forms a new form of research area that is different from the existing traditional wired networks. 

Regardless of which form of network is used, most of the data in networks is volatile, and volatility of data causes serious issues in forensic investigations. Most of the hardware used in networks record transmitted data itself or some other information about that data in logs. These logs are indispensable to the forensic investigators as they may contain information which can eventually be used as evidence.

Firewalls capture and record the information about network traffic and keep the logs of events and transmitted data which goes through them while preventing unauthorized access to the systems. \citet{jahanbin2013computer} proposed a design of autonomous intelligent multi-agent system in order to collect, examine and analyze firewall logs, and report possible evidence related to an ongoing or previous criminal activities in WSNs.

The proposed architecture is designed to be located between the firewall and the end user, and it consists of three cognitive agents.

\begin{itemize}
	\item \textit{The collector agent:} This agent is used in collection step and responsible for collection and processing of the firewall logs that are recorded for a given WSN.
	\item \textit{The inspector agent:} This agent is used in the inspection step and is responsible for identification of suspicious events from the given log files. It is also responsible for transmission of suspicious events to the next agent.
	\item \textit{The investigator agent:} This agent is used in both investigation and notification steps. In the investigation step, it examines the forwarded suspicious event by the inspector agent and evaluates its effects and importance. It eventually decides whether it is malicious or not. In the notification step, the decisions are reported as security alerts to the security administrator in details.
\end{itemize}

It must be noted that, in order to preserve forensic soundness, the firewall logs must be checked for integrity purposes as users (either an administrator or adversary) might alter the logs and destroy the evidence (intentionally or unintentionally). All the agents mentioned above work on the exact copy of the firewall log files and keep the originals as evidence in order to preserve integrity and provide reproducibility of forensic evidence.

Although WSNs have received the attention of security researchers, digital forensics research is still lacking in the discipline. In order to at least prepare WSNs for forensics investigations, \citet{mouton2011prototype} proposed a digital forensics readiness prototype in IEEE 802.15.4 WSNs. This prototype is designed based on the description made by \citet{tan2001forensic} who defines two digital forensics readiness objectives as:
\begin{enumerate}
	\item Maximizing an environment's ability to collect credible digital evidence,
	and;
	\item Minimizing the cost of forensics in an incident response.
\end{enumerate}

\begin{figure} [htb]
	\subfloat[A graphical representation of a wireless sensor network \citep{mouton2009secure}] {\includegraphics[width=0.49\textwidth,height=\textheight,keepaspectratio]{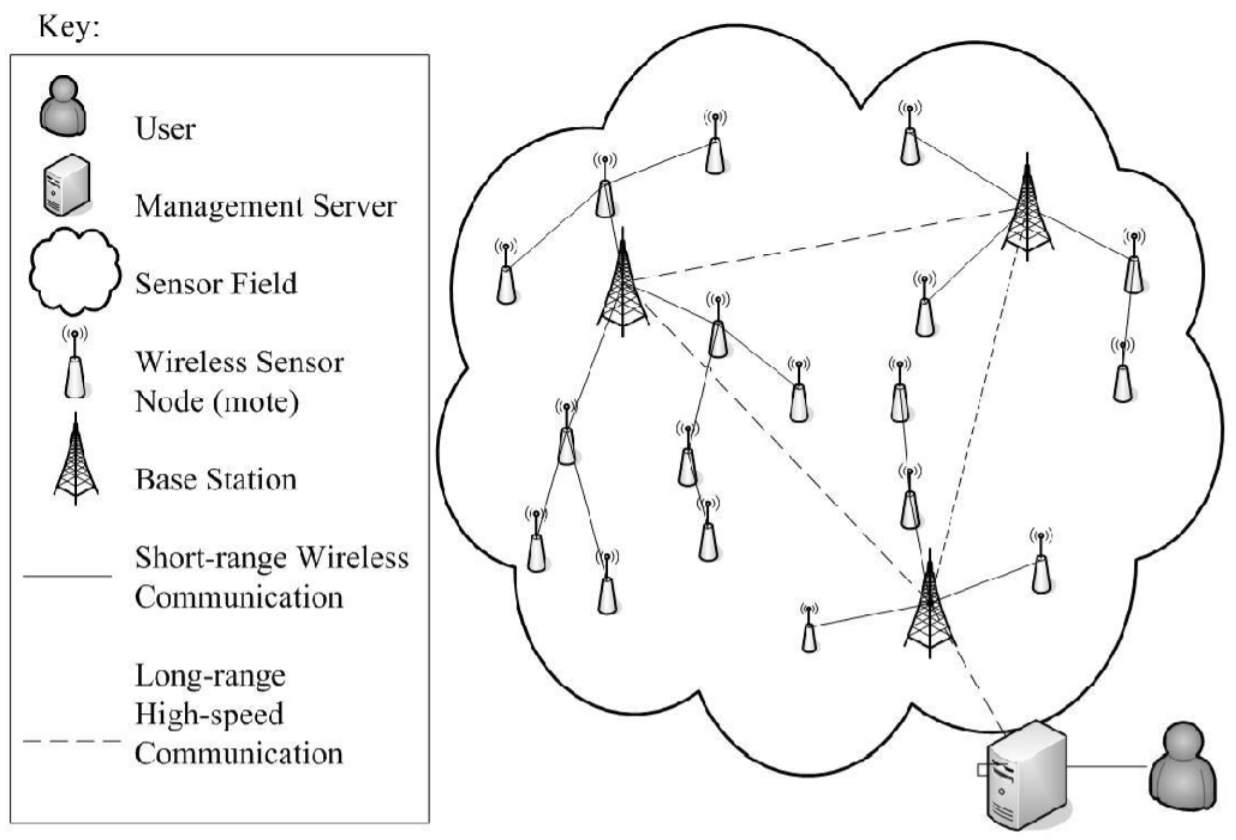}}
	~~~
	\subfloat[A graphical representation of the network layout with digital forensics readiness implemented \citep{mouton2011prototype}] {\includegraphics[width=0.48\textwidth,height=\textheight,keepaspectratio]{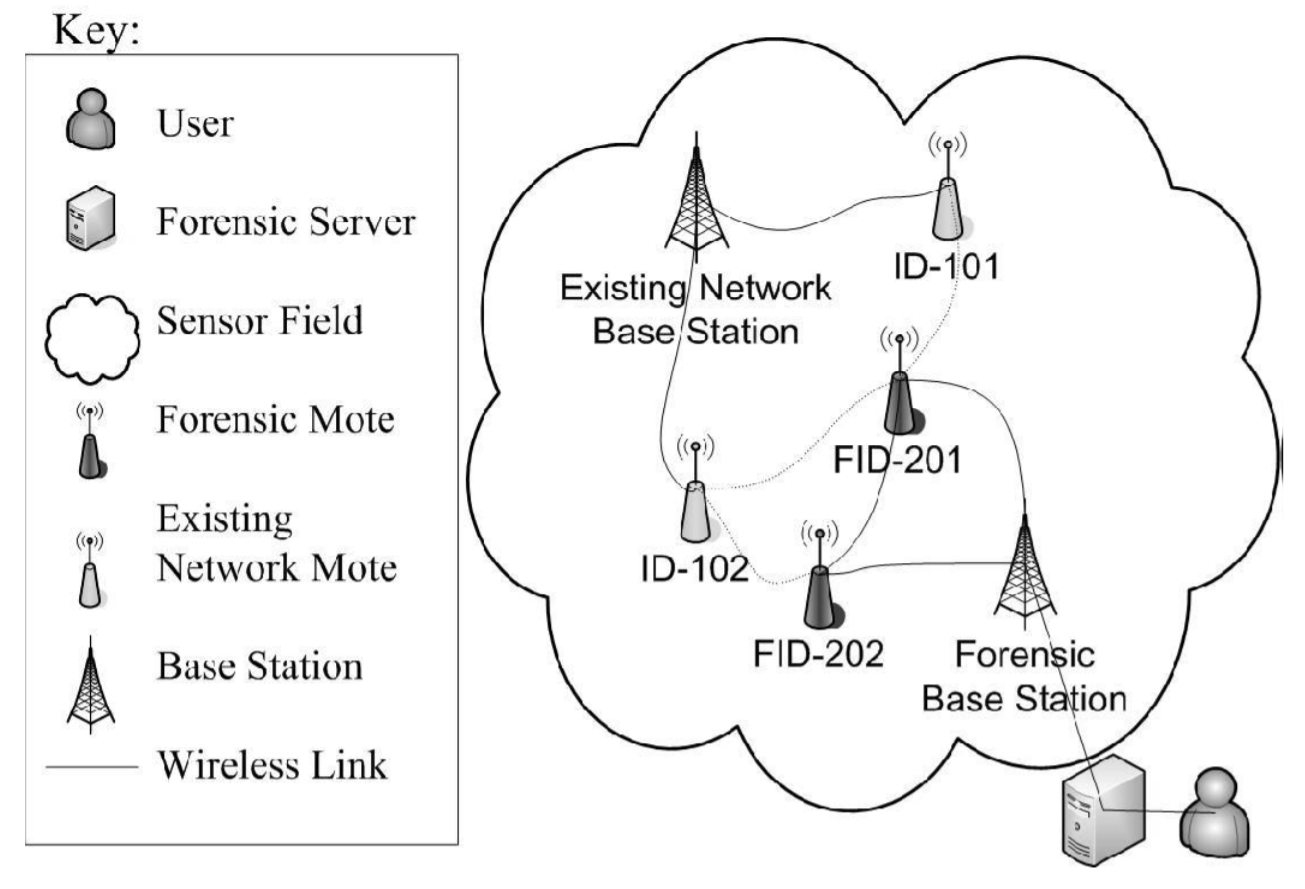}} 
	
	\caption{Adding a digital forensics readiness layer to an existing Wireless Sensor Network for digital evidence collection}
	\label{fig:wsngraph}
\end{figure}

Although \citet{tan2001forensic}'s objectives are sufficient enough for general digital forensics investigations, Mouton and Venter modified these objectives to be better suited to WSNs. Their objectives are threefold and aim to perform the investigation in the shortest amount of time, spending the least amount of time, and without causing disruptions in the network which may perform mission-critical tasks \citet{mouton2011prototype}. As digital forensics investigations require the original source of evidence being protected against alterations, the last objective is critical for forensic soundness. In order to avoid inadmissibility of evidence and make the implementation of digital forensics readiness, the authors created additional independent forensics WSN referred as \textit{fWSN} (see Fig. \ref{fig:wsngraph}(a)), along with the original WSN referred as \textit{oWSN} (see Fig. \ref{fig:wsngraph}(b)).

Mouton and Venter also discuss the list of requirements (see Table \ref{tab:dfreadiness}) which can be used as a preliminary approach and need to be considered in order to implement digital forensics readiness in an IEEE 802.15.4 WSN environment. Note that the first column in the table shows some important factors which make WNSs environments unique and different from WLAN. 

\begin{table}[H]
	\caption{Requirements in order to achieve digital forensic readiness in a IEEE
		802.15.4 WSN environment \citep{mouton2011requirements}}
	\label{tab:dfreadiness}   
	\setlength\extrarowheight{3pt}
	\begin{tabular}{p{4cm}p{0.3cm}p{8cm}}
		\hline
		\textbf{Factors} &	& \textbf{Detailed requirement list}\\
		
		\hline
		Communication Protocol
		&	&1. The \textit{fWSN} ensures the collection of all data packets by motes in the field utilizing a receipt acknowledgement packet protocol.\\
		&	&2. In order to make sure that the data packets are not changed, \textit{oWSN}'s broadcasting communication should be intercepted.\\
		&	&3. All p\textit{oWSN} ossible communication that is originating from \textit{oWSN}.\\
		
		\hline
		Proof of Authenticity and Integrity 
		&	&4. While \textit{fWSN} captures the data, the authenticity and integrity of all the data packets should be preserved.\\
		&	&5. Authenticity and integrity of the captured data in the \textit{fWSN} should be preserved while they are being stored.\\
		&	&6. Verification on the authenticity and integrity of all the data packets should be available when digital investigation takes place.\\
		
		\hline
		Time Stamping
		&	&7. The data packets should have a time stamp assigned to them in order to preserve their authenticity and integrity. \\
		&	&8. The order of the captured packets should reflect the correct sequence when compared to the data transmitted from the original network. \\
		
		\hline
		Modification of the network	after deployment
		&	&9. It should be possible to implement the \textit{fWSN} without any alteration in the \textit{oWSN}.\\
		
		\hline
		Protocol Data Packets
		&	&10. \textit{fWSN}’s operation should not be affected by the routing protocol or the network topology being used by \textit{oWSN}.\\
        		
		\hline	
		Radio Frequencies
		&	&11. The \textit{fWSN} should be able to communicate on the same radio frequencies that are available to the \textit{oWSN}.\\
		
		&	&12. All communication within the \textit{fWSN} should occur on a frequency not utilized in the \textit{oWSN}.\\
		&	&13. Data packet should be captured forensically by the \textit{fWSN} when an intruder WSN is in the area and communicates on a frequency that influences the \textit{oWSN}. \\
		
		\hline
		Power Supply
		&	&14. In order to ensure that the \textit{fWSN} captures all forensically relevant packet, the \textit{fWSN} should have at least the same or a longer network lifetime than the \textit{oWSN} in terms of battery power. Also, the \textit{fWSN} should not increase power consumption in the \textit{oWSN}.\\
		
		\hline					
		Network Overhead 
		&	&15. While intercepting communication, the \textit{oWSN} should be free of extra network overhead.\\
		
		\hline
		Data Integrity 
		&	&16. The \textit{fWSN} should by no means be able to influence the \textit{oWSN} or influence any sensory data transmitted within the \textit{oWSN}.\\
		
		\hline
		
	\end{tabular}
\end{table}

\begin{figure}[htb]
	\centering
	\includegraphics[width=0.6\textwidth,height=\textheight,keepaspectratio]{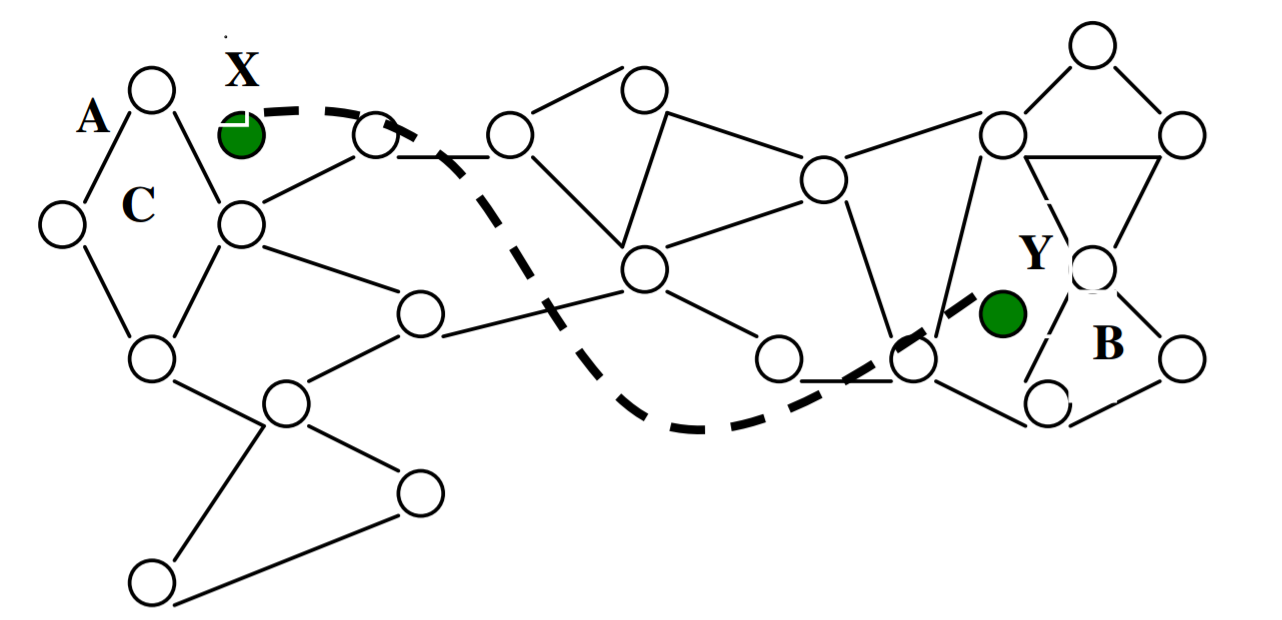}	\caption{Wormhole Attack~\citep{arora2010performance}}
	\label{fig:wormhole}
\end{figure}

In another work, \citep{triki2009digital} propose a solution to digital forensics investigations when wormhole attacks take place in a WSN. This solution ultimately aims to collect digital evidence, detect colluded nodes and reconstruct the events which occurred during the wormhole attack which allows attackers to transmit a packet from one point to another point in the network by creating `tunnels'  (see Fig. \ref{fig:wormhole}). This will allow attackers to distribute the packet to other nodes from the second point in the network (see \citet{arora2010performance} for attack details). 

\begin{figure}[htb]
	\centering
	\includegraphics[width=0.7\textwidth,height=\textheight,keepaspectratio]{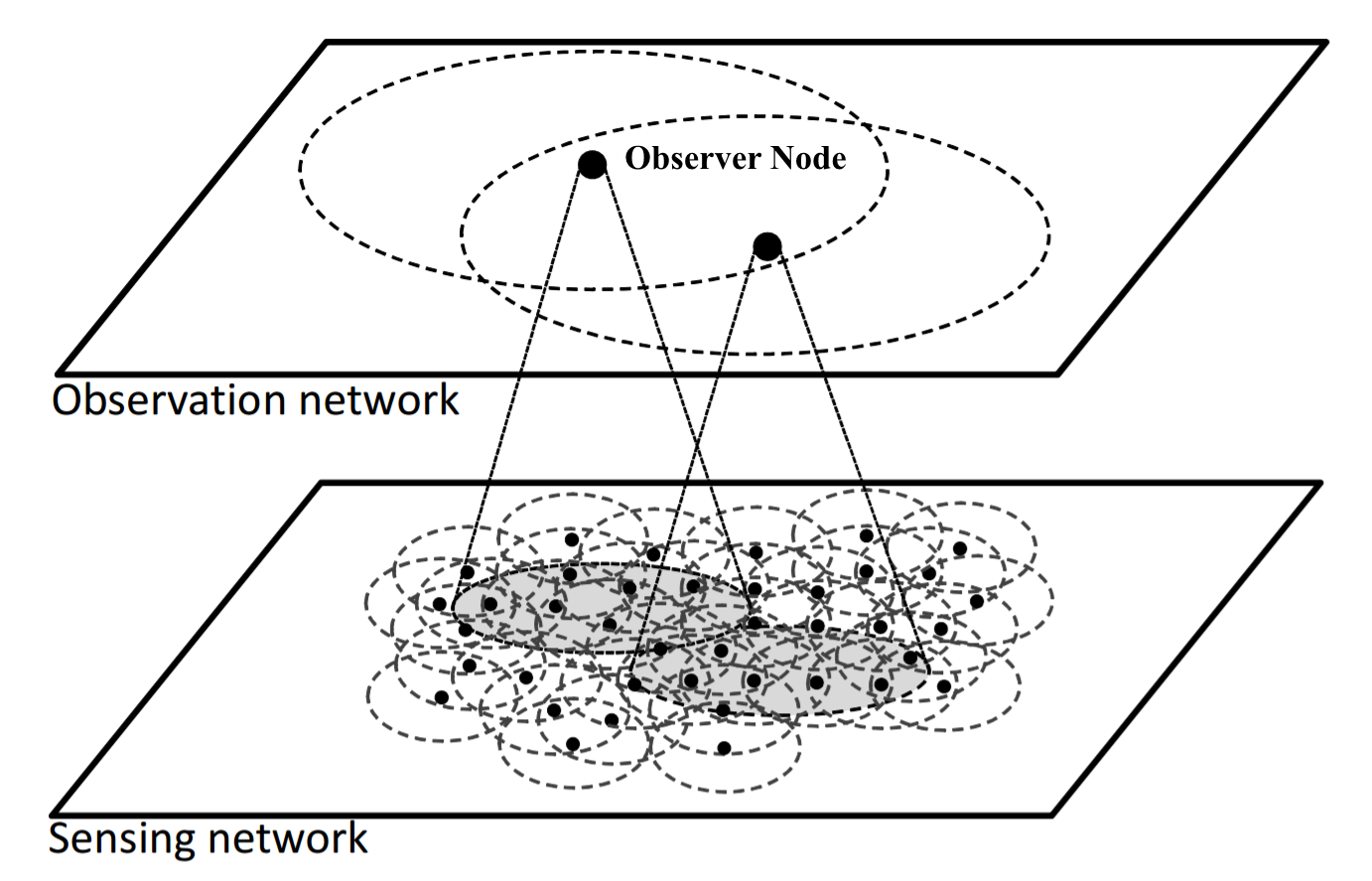}
	\caption{Proposed WSN architecture \citep{triki2009digital}}
	\label{fig:observer}
\end{figure}

The proposed solution suggests creating a virtual network called observation network which consists of a set of investigator nodes and base stations. The nodes in this secondary network are called \textit{observers}. Each observer has a limited coverage and they are responsible for monitoring the communication between observed nodes located in the sensing network (see architecture in Fig.~\ref{fig:observer}). \textit{Observers} collect information about the suspicious nodes such as traffic between nodes, routing path of data packets, and identity of those nodes. The aggregated evidence data is then broadcast to base-stations.

Base stations are responsible for several activities which are defined in the following two groups \citep{triki2009digital}.
\begin{enumerate}
	\item Sensing based activities such as:
	\begin{itemize}
		\item Analyzing collected data from sensors
		\item Creating decisions by correlating and filtering the data
		\item Transmitting configuration to sensors
		\item Activating sensors dynamically to reduce battery usage
	\end{itemize}
	
	\item Investigation based activities such as:
	\begin{itemize}
		\item Collecting forensic information about the observed nodes from observation network 
		\item Analyzing, correlating and merging the evidence in order to determine malicious nodes and rebuild the attack scenarios
		\item Communicating with observers about configuration of their locations
	\end{itemize}
\end{enumerate}

This proposed architecture is able to detect all types of wormhole attacks as the observer nodes and the observation network is designed in such a way that no observed nodes are left behind in the sensing network. In other words, all the observed nodes are clustered into groups and each cluster is constantly observed by observer nodes.\\

\subsubsection{CyberForensics for CPS and SCADA.}
Recently, IoT has also started to be deployed in control systems for actuation purposes which led to the concept of CPS \citep{rajkumar2010cyber,khaitan2015design}. In such systems, IoT devices are involved in sensing, communicating and acting. The difference from the above networks is that there are nodes which do actuation and thus this creates another venue for forensics investigation. One form of CPS is in the area of control systems for critical applications such as energy, transportation and industry. In such systems, the network is referred to as Supervisory Control and Data Acquisition (SCADA) \citep{boyer2009scada} and failure or attacks in such systems is crucial to be detected and investigated for the applications to sustain \citep{krutz2005securing}. SCADA systems, are used for the collection and analysis of real-time data from Industrial Control Systems (ICS). Most of the CPSes rely on computer and control systems in order to provide reliable operations to safeguard the infrastructure. Therefore, forensics analysis of SCADA/ICS systems has been an important tool which was considered in some works. In the remainder of this section, we also discuss these approaches as they relate to a network-level investigation.  

SCADA systems consists of a field site and control center. In the field site, there are IoT devices which are considered as intelligent such as Programmable Logic Controllers (PLCs) or Remote Terminal Units (RSUs). These are typically attached to physical processes such as thermostats, motors, and switches. The control center is responsible for collecting data related to the state of field instruments and interacting with the field sites. Components found at the control center typically consist of a Human Machine Interface (HMI), Historian and Master Terminal Unit (MTU). All of these are connected with a LAN that can run various protocols including MODBUS \cite{modbus2004modbus}, DNP3 \cite{clarke2004practical}, and Ethernet. 

The information security vulnerabilities of ICS have been studied extensively, and the vulnerable nature of these systems is well-known \citep{stouffer2011guide,patzlaff2013d7, shahzad2014industrial}. However, in the case of a security incidents (e.g., denial of service attacks), it is important to understand what the digital forensics consequences of such an attack are? What procedures or protocols are needed to be used during an investigation? What tools and techniques are appropriate to be used by the investigator? Where can forensic data be collected and how? 

In the rest of this section, we discuss various research efforts aimed at assisting SCADA forensics procedures by proposing tools, techniques and forensics investigation models.

\paragraph{SCADA Live Forensics:} SCADA is originally deployed to non-networked environments, therefore there has been a lack of security against Internet-based threats and cyber-related forensics. Over time, there has been a huge increase in the vulnerability of threats caused through connectivity allowing remote control over the Internet. The attacks necessitated SCADA system a forensic investigation in order to understand the effects and cause of the intrusion. \cite{taveras2013scada} focuses on detecting the abnormal changes of sensor reads, illegal penetrations, failures, traffic over the communication channel and physical memory content by creating a software application. The challenging issue is that the tool should be developed in a way that it should have the minimal impact over the SCADA resources during the data acquisition process.

The problem involved in this process is that SCADA systems should not be turned off for data acquisition and analysis as it is being continuously operational. There has not been a single forensic tool to preserve the hardware and software state of a system during investigation. Research continued to provide a computing module to support the incident response and digital evidence collection process. Experiment is performed on the SCADA system by performing live data acquisition and then performing subsequent offline analysis of the acquired data. 

Based on the live forensic analysis of the data collected from the SCADA system, it is concluded that traditional information security mechanisms cannot be applied directly as these systems cannot tolerate delays in performance which eventually require a lot of memory to perform long processes. Thus, SCADA systems should consider a special operating paradigm. This also paved the way to improve the infrastructure of the systems and provide appropriate tools for forensic analysis over interconnected SCADA systems.

\paragraph{Limitations of forensic analysis tools on SCADA systems:} As \cite{ahmed2012scada} notes, currently available traditional digital forensics tools are not capable of performing data analysis on SCADA systems. The main reason is that state-of-the-art tools are designed to work on deterministic systems and devices such as hard disk drives, mobile phones, network traffic captures saved in pcap files. However, SCADA systems generate propriety log data depending on the make and model the hardware. As discussed above, investigators are in need of creating new scripts for their own particular needs to overcome this issue. Hence, there is an expectation from the research community and forensics tools developers to design SCADA forensics tools or patch currently available tools in order to respond to this demand.   

\paragraph{Developing Forensics Investigation Models for SCADA:} Once the vulnerabilities and the possible attacks on the SCADA/ICS systems are analyzed, it is crucial to perform forensically sound forensic analysis on SCADA/ICS. The current literature shows some efforts of developing forensic analysis frameworks and models. 

\begin{figure}[htb]
	\centering
	\includegraphics[width=0.85\textwidth ,keepaspectratio]{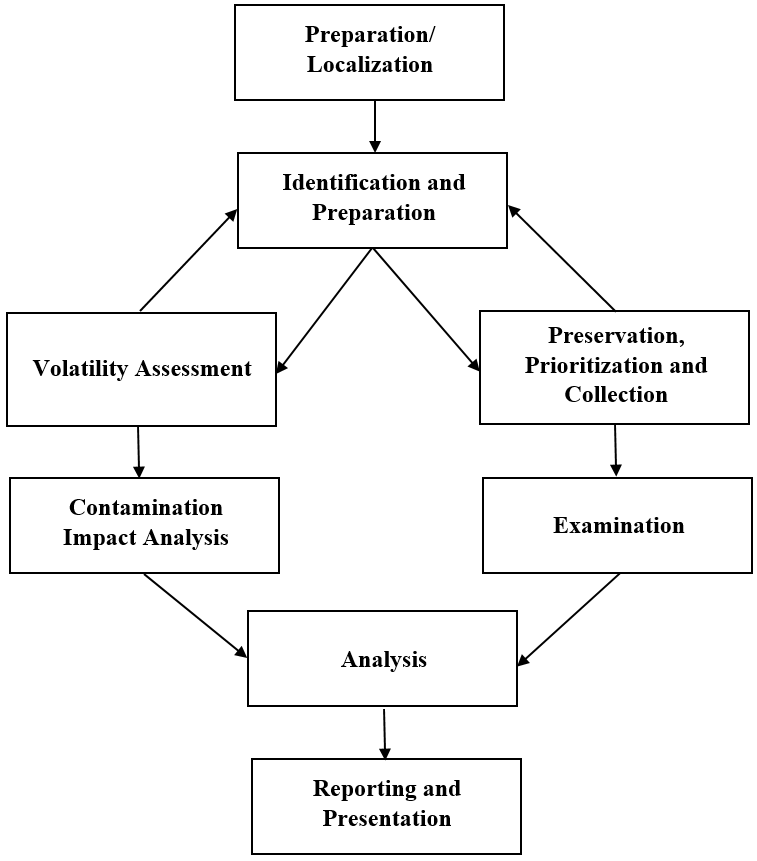}
	\caption{Cyber Forensics Model for SCADA/ICS by \cite{stirland2014developing}}
	\label{fig:SCADA_model}
\end{figure}

One of the early frameworks is proposed by \cite{wu2013towards}. In addition to this framework, \cite{stirland2014developing} proposed a methodology to analyze the problems involved in SCADA/ICS systems proposed (see Fig. \ref{fig:SCADA_model}). The authors in this work particularly categorize a set of forensic toolkits (both commercial and open source) to support each stage of an investigation and structure of  the control systems. The proposed methodology involves a clear process of investigation which includes the following phases: 

\begin{table}[tb]
\caption{Forensic Toolkit Application to SCADA Systems}
\label{tab:dftools}   
\centering
\setlength\extrarowheight{3.5pt}
\begin{tabular}{|c|c|c|}
\hline
\textbf{SCADA Device}                                                                    & \textbf{Phase} & \textbf{Forensic Tool}                                                                                         \\ \hline
\multirow{2}{*}{\textbf{Network}}                                                        & Phase 3        & TCPDump                                                                                                        \\ \cline{2-3} 
                                                                                         & Phase 4        & Network Miner, Wireshark, AlienVault                                                                           \\ \hline
\multirow{2}{*}{\textbf{HMI}}                                                            & Phase 3        & \begin{tabular}[c]{@{}c@{}}Write Blockers, FTK Imager, EnCase, Helix, \\ SHA-256/MD5 Hashing Tool\end{tabular} \\ \cline{2-3} 
                                                                                         & Phase 4        & EnCase, XWays, Accessdata FTK, Volatility                                                                      \\ \hline
\multirow{2}{*}{\textbf{PLC/RTU}}                                                        & Phase 3        & Besope PLC Flashing Software                                                                                   \\ \cline{2-3} 
                                                                                         & Phase 4        & XWays                                                                                                          \\ \hline
\multirow{2}{*}{\textbf{\begin{tabular}[c]{@{}c@{}}Engineering\\ Computer\end{tabular}}} & Phase 3        & \begin{tabular}[c]{@{}c@{}}Write Blockers, FTK Imager, EnCase, Helix, \\ SHA-256/MD5 Hashing Tool\end{tabular} \\ \cline{2-3} 
                                                                                         & Phase 4        & EnCase, XWays, Accessdata FTK, Volatility                                                                      \\ \hline
\multirow{2}{*}{\textbf{\begin{tabular}[c]{@{}c@{}}Database\\ Server\end{tabular}}}      & Phase 3        & \begin{tabular}[c]{@{}c@{}}Write Blockers, FTK Imager, EnCase, Helix, \\ SHA-256/MD5 Hashing Tool\end{tabular} \\ \cline{2-3} 
                                                                                         & Phase 4        & EnCase, XWays, Accessdata FTK, Volatility                                                                      \\ \hline
\multirow{2}{*}{\textbf{OPC}}                                                            & Phase 3        & \begin{tabular}[c]{@{}c@{}}Write Blockers, FTK Imager, EnCase, Helix, \\ SHA-256/MD5 Hashing Tool\end{tabular} \\ \cline{2-3} 
                                                                                         & Phase 4        & EnCase, XWays, Accessdata FTK, Volatility                                                                      \\ \hline
\multirow{2}{*}{\textbf{Historian}}                                                      & Phase 3        & \begin{tabular}[c]{@{}c@{}}Write Blockers, FTK Imager, EnCase, Helix, \\ SHA-256/MD5 Hashing Tool\end{tabular} \\ \cline{2-3} 
                                                                                         & Phase 4        & EnCase, XWays, Accessdata FTK, Volatility                                                                      \\ \hline
\end{tabular}
\end{table}

\begin{enumerate}
\item Identification and preparation of the requirements and the problem involved in extracting the evidence.
\item Identifying data sources- this phase involves gathering the data from sources and analyzing if the system supports the data sources.
\item Preservation, prioritizing and collection – this phase works depending on the priority of data and different data capturing techniques are involved to ensure all devices are captured or not.
\item Examination and analysis- this phase involves in performing the analysis depending on the data sources, methods and provides a timeline in preparing the data and logs on it and allows to extract data.
\item Reporting and presentation- this phase involves in providing a report to all the details performed in the above phases including the outcome of the analysis which also includes documentation of the further recommendations for future study.
\end{enumerate}

Security is of high importance for the control systems and there are many recommendations for further improvements in incident response to support investigation and increase the level of complexity for attacking the systems by attackers. It is concluded that the proposed methodology for developing a forensics toolkit is considered based on the requirements of SCADA systems. Various suggested tools are shown in Table \ref{tab:dftools}. There are already existing tools which support the elements of SCADA forensic investigation and further research and progress in this area is needed in order to identify more evidence and artifacts.

\begin{figure}[H]
	\centering
	\includegraphics[width=0.7\textwidth ,keepaspectratio]{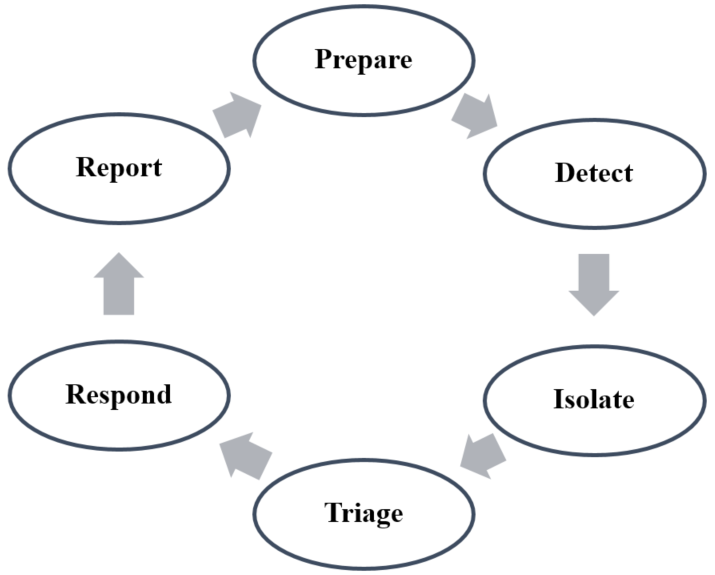}
	\caption{Forensics Model for SCADA Systems}
	\label{fig:SCADA}
\end{figure}

Moreover, \cite{ahmed2012scada} discuss the challenges for forensics investigators in SCADA systems and their potential solutions. In order to address some of these unique challenges of SCADA forensics analysis, a recent framework was proposed in \citep{Eden2016}. This framework aims at identifying necessary steps for incident response as well. Fig. \ref{fig:SCADA} shows the SCADA forensic incident response model consisting of six main stages:  1) Prepare; 2) Detect; 3) Isolate; 4) Triage; 5) Respond; and 6) Report.

This proposed model suggests that the preparation phase must be performed prior to incident the happening. In this first stage, an investigator must understand the system's architecture with respect to its configuration and hardware devices used in the system. This step is crucial to the first responders to avoid complication when the recovery from an incident is time sensitive. In this stage, it is made sure that all the hardware used in the system is well documented. In addition to the system architecture, the forensics investigator is also expected to know the SCADA system's requirements with respect to availability of the system. This is essential particularly if the system is mission critical and running states of a certain device must be preserved while the investigation is in progress. Therefore, prior knowledge of system requirements play a critical role. Finally, the investigator is also expected to understand potential attacks targeting hardware, software and the communication stack of the SCADA system. Such knowledge will allow the investigators to better perform in the following phases. 

In the second stage of the model, the investigator is expected to detect the type of attack and potential infected areas in the system. This detection will be performed based on the real-time data available in the system such as network packages and log files. Once the type of attack has been determined, investigators will be able to locate infected areas based on the behavior of the attack. As long as the infected areas are detected, it will be easier to know the type of data in the next stage.

The isolate stage is critical for the investigation with respect to the importance of SCADA system in the CPS. In most of cases, infected areas in the network must be isolated so that further contamination and disruption to the system can be avoided. The success of the isolation will be dependent on the success of the detection of potential infected areas. 

Despite the classical forensics investigations, triaging is different in SCADA networks. Forensics investigators must identify the data sources in order to triage the available data. This will also be dependent on the information (e.g., device make, model, and serial number) provided in earlier stages, particularly preparation and detection stages. Once the data identification is performed, then data sources ought to be prioritized with respect to the value, volatility and accessibility of data. This will allow investigators to acquire as much evidence as possible.

In the respond stage, investigators perform actual data acquisition from the SCADA system (network) by using the priority list created in the previous stage. As a rule of thumb, data must be acquired from the SCADA system by using forensically sound tools and techniques. As discussed before, this will prove the admissibility of evidence in a court of law (when needed). In order to acquire data from various devices and network, aforementioned forensics acquisition techniques can be used. Once the data acquisition is completed, then analysis of data is performed using available forensics tools or by creating new special scripts for unconventional data. Eventually, the aim is to find a forensics artifact to be presented as evidence from the large set of unrelated system data.

Finally, similar to the traditional investigations, the reporting stage requires investigators to document all the steps taken, tools used, evidence collected, and challenges faced. When they are documented systematically, then the investigator may create a timeline of events by reviewing the findings to support the evidence found and determine the source of an incident/attack. The final report must comply with the chain of custody by providing validation and verification of evidence found.

Our final discussion in this section is briefly on accurate modeling of the SIEMENS S7 SCADA Protocol for intrusion detection and digital forensics using real-life data. Siemens S7 is used in SCADA systems for communications between a HMI and the Programmable Logic Controllers (PLCs). In \citep{kleinmann2014accurate}, Intrusion Detection system (IDS) model is designed for S7 networks which analyzes the traffic to and from a specific PLC. A unique Deterministic Finite Automata (DFA) is used to model the HMI-PLC channel traffic whether it is highly periodic or not. 

SCADA systems have its own strategy in analyzing the fault or malfunction. In this paper,  it is defined that the research based on traffic simulation has several risks such as lack of realism which effects the use of SCADA systems. Three different traces of datasets are collected in order to perform the experiments which are collected at ICS facilities. The first S7 SCADA trace was collected from a manufacturing plant where a single channel is observed between the HMI and an S7 complaint VIPA PLC. Next two traces are collected from a water treatment facility which has control over specific levels in tanks. A Wireshark program is used to collect the traces with HMI running in background in the operating system.
Authors show that, based on the analysis of the traffic from two ICS plants, some key semantics of the proprietary of S7 protocol can be reverse engineered. It is also observed that previously developed Modbus showed successful results in the same way DFA-based approach is very successful with high accuracy and extremely low-false positive rates; IDs is further extremely efficient which works at line speed to detect the anomalies.

\subsection{Cloud Level Investigation}\label{cloudLevel}

As discussed in the previous sections, forensic investigation in cloud environments has its own challenges such as multi-tenancy and multi-jurisdiction. Since IoT devices have limited storage and computational resources, the actual data is processed and stored in the cloud. This causes investigations being conducted in the cloud environment especially when data in physical storage and network does not result useful evidence. Hence, similar investigative challenges in the cloud exist when forensic investigations in IoT are conducted. Although current research efforts in IoT forensics are in their very early stages, there are some successful models suggesting easier investigations in the cloud environment. In this section we will specifically focus on IoT forensics investigation models proposed for cloud environments.

\begin{figure}[H]
	\centering
	\includegraphics[height=0.8\textheight,keepaspectratio]{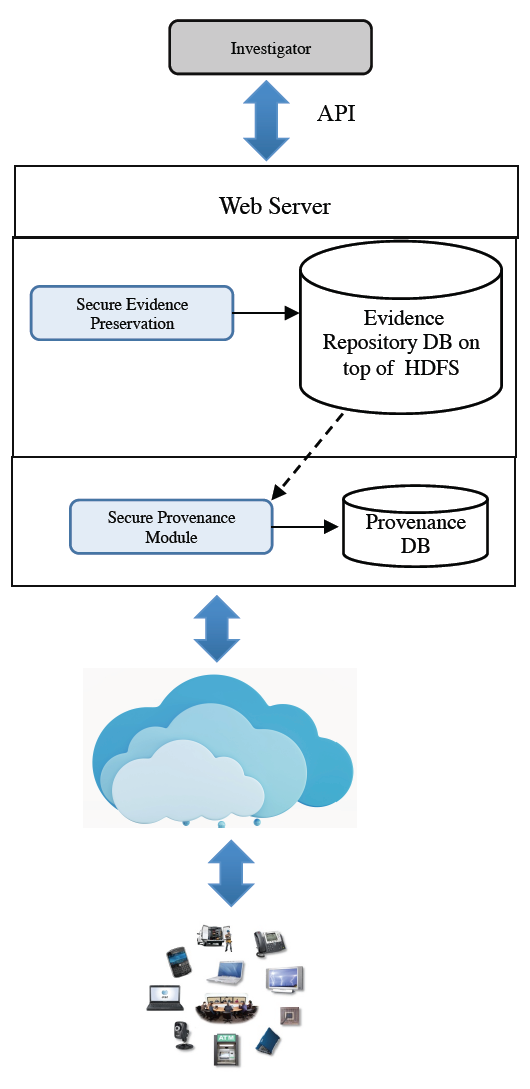}
	\caption{Proposed Model of IoT Forensics \citep{zawoad2015faiot}}
	\label{fig:faiot}
\end{figure}

According to \citet{zawoad2015faiot}, the term IoT forensics was not formally defined until they proposed forensics-aware model (see Fig. \ref{fig:faiot}) for IoT infrastructures called FAIoT. This model supports digital evidence collection and analysis in the IoT environment by providing easiness and forensic soundness. Such a model might also allow cloud service providers addressing the needs of law enforcement officers when a search warrant is obtained to collect data from cloud environments.

\paragraph{Secure Evidence Preservation Module:} This module provides constant monitoring of the registered devices for forensics evidence in the form of logs files or data collected by sensors. If evidence is recognized, it is then stored in the evidence repository. Evidence repository database is designed on top of Hadoop Distributed File System (HDFS) in order to provide scalable and reliable data processing of large data. The data kept in the database will be categorized based on the IoT device and its owner in order to reduce multi-tenancy issues and avoid commingling of data in cloud. \citep{zawoad2015faiot}.  

\paragraph{Secure Provenance Module:} This module provides chain of custody for the evidence stored and kept in the database. This is made possible by using provenance aware file system (PASS) introduced by \citet{muniswamy2006provenance}. PASS is a storage system which performs automated collection, management, storage and search of provenance an object \citep{muniswamy2006provenance}. Secure Provenance Module provides provenance record of evidence stored in provenance database by using PASS.

Finally, investigators can access the evidence and its provenance record using proposed APIs which makes sure the confidentiality of evidence by using encryption algorithms. In order for this to be possible, investigators need a Web Server to access the requested data through the API.

\begin{figure}[H]
	\centering
	\includegraphics[width=\textwidth, height=\textheight,keepaspectratio]{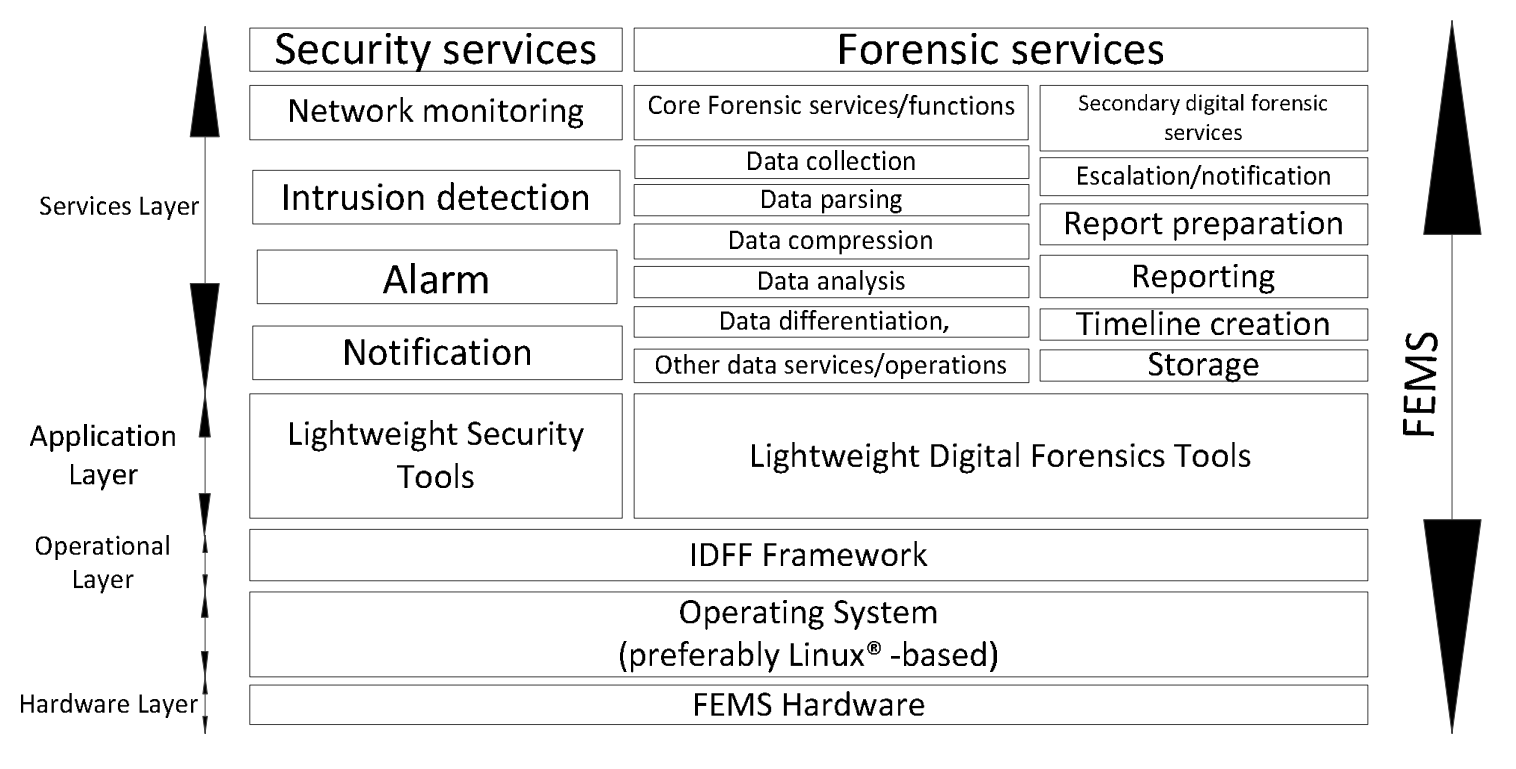}
	\caption{The FEMS Architecture \citep{oriwoh2013forensics}}
	\label{fig:fems}
\end{figure}

In another work, \citet{oriwoh2013forensics} propose a more specialized model called Forensics Edge Management System (FEMS) which is specific to smart home environments. FEMS is an automated system which can be integrated into smart homes in order to perform initial forensic investigations while providing basic security services \citep{oriwoh2013forensics}. Fig. \ref{fig:fems} shows the architecture of the FEMS and all the security and forensic services provided by FEMS.

\begin{figure}[H]
	\centering
	\includegraphics[width=\textwidth, height=\textheight,keepaspectratio]{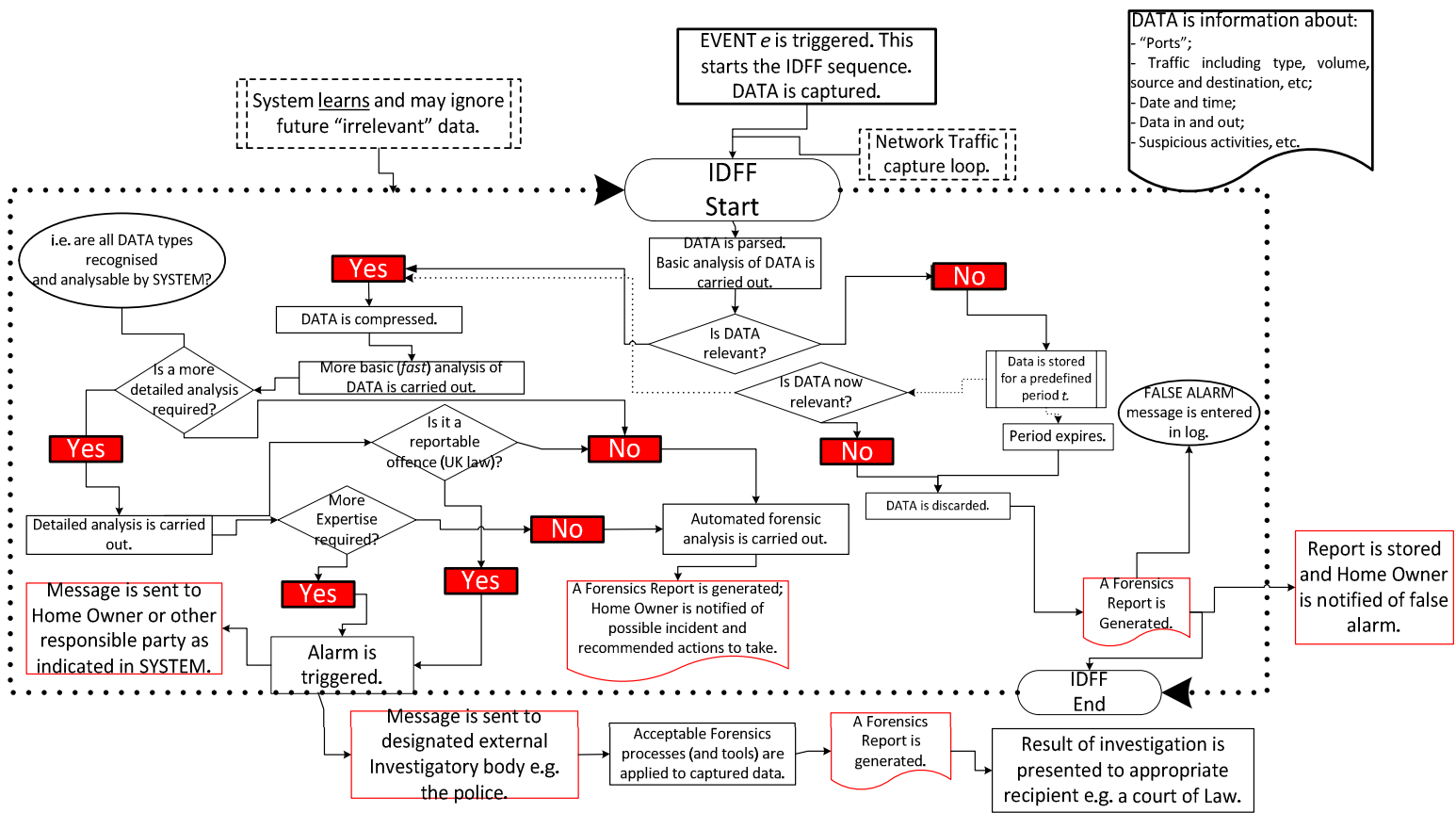}
	\caption{ Detailed flowchart of the IoT Digital Forensics Framework (IDFF) \citep{oriwoh2013forensics}}
	\label{fig:idff}
\end{figure}

Oriwoh and Sant also proposed a digital forensics framework called IoT Digital Forensics Framework (IDFF) (see Fig. \ref{fig:idff}). This framework presents step by step operation presented in the flowchart in order to show how FEMS operation is performed. As stated by the authors, usage of FEMS provides automatic, intelligent, and autonomous detection and investigation, and indicates the source of security issues in smart homes to its users.

\subsection{Future Research} \label{sec:future_research}
As digital forensics in IoT and WSNs is a relatively new concept particularly for the digital forensics community, the current research reveals important future work to be conducted. It is quite obvious that newer investigative techniques will be soon needed as Google has announced Android Things \citep{androidThings} a new operating system to develop new IoT devices and there is a growing number of IoT devices being deployed in our daily applications. It is worth noting that as WSNs are now widely considered under the IoT concept, most of the endeavors for digital forensics research will be also applicable for WSNs. In addition, digital forensic solutions and frameworks also will be needed due to the availability of low memory footprint and low power requirement devices used in WSNs.

Despite this inevitable demand in the very near future,  once they are developed, these new methods will ultimately help investigators to perform standard investigative processes. Until then, it could also be possible to use some of the available tools and techniques that are readily available for current Android operating systems. In addition to the existing research literature, we believe some of the areas for further work could be listed as follows:

\begin{itemize}
	\item standardize data storage units and interfaces in similar devices. Forensically valuable IoT devices (e.g. fitness trackers) could be designed and manufactured with data storage units which can be analyzed using state of the art forensic tools. Using known interfaces such as JTAG connections for IoT devices is also critical for faster and reliable investigations.
    \item develop automated decision-making systems on forensically sound data for specific IoT technologies such as smart homes. It is well known that artificial intelligence techniques have been applied to many digital forensic domains to intelligently automate duties performed by human entities. Therefore, as an example, it can be very useful to adapt machine learning techniques to classify evidence in IoT domain or expert systems can be used to create intelligent tools to make decisions based on knowledge collected from both investigators and IoT environments.
    \item build a model that would correlate evidence found in IoT environments. Digital forensics evidence correlation is an important concept especially when heterogeneous data is involved in investigations. \citet{case2008face} have developed a framework for automatic evidence discovery and correlation from a variety of forensic targets. We also believe that similar models can also be built for IoT environments in order to use unrelated data leading to  actual evidence through correlation.  
    \item analyze Android Things and develop new forensics models and tools for data acquisition, examination, analysis and reporting. This brand new operating system needs immediate attention from the researchers as it is projected to be used in many IoT devices in the near future. 
    \item create new digital forensics investigation models (e.g. Electronic Discovery Reference Model, see http://www.edrm.net) for specific IoT environments. Due to the heterogeneity of data and hardware in IoT devices, it could be useful to develop IoT specific investigation models. Because, currently available models are mostly designed for storage, network, and cloud specific, however, IoT environments may necessitate all three environments being used.
    \item collaborate with data analytics and fault-tolerance experts to cooperatively analyze data from IoT devices not only related to user activity but also related to hardware and embedded systems. This opens up opportunities for insurance companies as they would like to investigate issues regarding failures while some of these failures might be due to actual attacks from external attackers. 
    \item create robust and standard solutions particularly for live data acquisition, automated data collection, recovery of memory and processes from live units in SCADA systems. 
    \item develop legal solutions to the issues including preservation of the chain of custody and admissibility of IoT evidence. In digital investigations, it is critical to preserve chain of custody for evidence admissibility. However, it may not be possible in IoT environments because of their designs. Involvement in legislative processes regarding IoT forensics investigations is needed to determine solutions from the legal aspects.
\end{itemize}

\subsection{Conclusion}
The IoT and WSNs offer a significant source of potential evidence, however due to their heterogeneous nature and the ways in which data is distributed, aggregated, and processed, there are challenges that the digital forensics investigations must overcome. For this purpose, new techniques are required to not only overcome the hurdles, but also influence the architecture and processes in order to gain access to this rich source of potential evidence in the IoT and thus WSN environments. In this book chapter, we explained digital forensics challenges in IoT and WSN environments. We also analyzed and explained currently available solutions to overcome some of those challenges from different perspectives. As discussed in the Section \ref{sec:future_research}, there are still many open research problems in this new area.

\bibliography{references}

\end{document}